\documentclass[preprint,10pt]{aastex}
\usepackage{amssymb}
\usepackage{natbib}
\usepackage{amsmath}
\usepackage[usenames,dvipsnames]{color}
\usepackage{ulem}
\usepackage{graphicx}
\usepackage{morefloats}

\begin{document}

\title{Spectral Energy Distributions of Young Stars in IC~348: 
The Role of Disks in Angular Momentum Evolution of Young, Low-Mass Stars}

\author{Thompson S.\ Le Blanc\altaffilmark{1,2}, 
Kevin R.\ Covey\altaffilmark{3,4}, 
Keivan G.\ Stassun\altaffilmark{1,5}}

\altaffiltext{1}{Department of Physics \& Astronomy, Vanderbilt University, 6301 Stevenson Center, Nashville, TN, 37235}
\altaffiltext{2}{NASA Graduate Student Research Fellow}
\altaffiltext{3}{Cornell University, Department of Astronomy, 226 Space Sciences Building, Ithaca, NY 14853}
\altaffiltext{4}{Hubble Fellow} 
\altaffiltext{5}{Department of Physics, Fisk University, 1000 17$^{th}$ Ave. N., Nashville, TN, 37208} 

\begin{abstract}
Theoretical work suggests that a young star's angular momentum content and
rotation rate may be strongly influenced by magnetic interactions with
its circumstellar disk.  A generic prediction of these `disk-locking'
theories is that a disk-locked star will be forced to co-rotate with
the Keplerian angular velocity of the inner edge of the disk; that is,
the disk's inner truncation radius should equal its co-rotation radius.
These theories have also been interpreted to suggest a gross correlation
between young stars' rotation periods and the structural properties of
their circumstellar disks, such that slowly rotating stars possess close-in
disks that enforce the star's slow rotation, whereas rapidly rotating stars
possess anemic or evacuated inner disks that are unable to brake the stars
and instead the stars spin up as they contract.  To test these expectations,
we model the spectral energy distributions of 33
young stars in IC~348 with known rotation periods and infrared excesses
indicating the presence of circumstellar disks.  For each star, we match the
observed spectral energy distribution, typically sampling 0.6--8.0 $\mu$m,
to a grid of 200,000 pre-computed star$+$disk radiative transfer models,
from which we infer the disk's inner-truncation radius.  We then compare
this truncation radius to the disk's co-rotation radius, calculated from
the star's measured rotation period.  We do not find obvious differences in
the disk truncation radii of slow rotators vs.\ rapid rotators. This holds
true both at the level of whether close-in disk material is present at all,
and in analyzing the precise location of the inner disk edge relative to the
co-rotation radius amongst the subset of stars with close-in disk material.
One interpretation is that disk-locking is unimportant for the IC~348 
stars in our sample. Alternatively, if disk-locking does operate, then it 
must operate on both the slow and rapid rotators, potentially 
producing both spin-up and spin-down torques, and the transition from
the disk-locked state to the disk-released state must occur more rapidly
than the stellar contraction timescale.
\end{abstract}


\section{Introduction \label{intro}}
Stars are born from the gravitational collapse of dense cores within giant
molecular clouds.  Conservation of angular momentum would imply that this
collapse should produce a star spinning at or near breakup velocity. Actual
observations, however, show that young stars rotate much slower than
breakup \citep[$\leqslant$ 10\% v$_{breakup}$,][]{Hartmann:1986aa,
Bouvier:1986aa}. Where the rest of the angular momentum goes is an important
outstanding question.

Circumstellar disks are one important angular momentum reservoir for young
stars.  During the process of protostellar collapse, a dense core's highest
specific angular momentum material forms a flattened disk which persists
for $\sim$2-3 Myr \citep[e.g.][]{Lada:2006aa} to $\sim$6 Myr \citep[e.g.][]{Haisch:2001aa}, before it is
depleted by accretion onto the star \citep{Bertout:1988aa, Hartigan:1995aa},
planet formation \citep{Mordasini:2009aa, Mordasini:2009ab}, outflows
\citep{Reipurth:1999aa, Hartigan:2005aa}, or photoevaporation
\citep{Bertoldi:1989aa}.

Theory suggests that magnetic star-disk interactions could play an
important role in the star's early angular momentum evolution
\citep{Konigl:1991aa,Shu:1994aa,Hartmann:2001aa, Hartmann:2002aa}.
\citet{Ghosh:1979aa, Ghosh:1979ab} and \citet{Konigl:1991aa} provided the
first analytic descriptions of this process, assuming steady state accretion 
and a star with a strong magnetic field. In this `disk-locking' picture,
angular momentum is transferred from a star to its circumstellar disk
via torques arising from interactions between the star's magnetic field
and ionized gas in its circumstellar disk.  As the star's magnetic field
weakens with distance, the field couples most strongly to the inner edge
of the disk, ``locking" the star's rotation to the Keplerian orbital period
at the inner edge of the disk. \citet{Shu:1994aa} extended this theoretical
framework by developing the `X-wind' model, in which a magnetically driven
wind carries angular momentum away from the ``X-point", where the star and
disk's magnetic fields pinch at the disk's co-rotation radius. 
This model, originally assuming a dipolar configuration for the star's
magnetic field, has been generalized by \citet{Mohanty:2008aa} to
include more complex field geometries.

\citet{Edwards:1993aa} and \citet{Edwards:1994aa}
provided some of the first observational evidence in support of the
disk-locking picture.  Analyzing rotation periods measured from stellar
light curves, these studies found that stars possessing close-in
circumstellar disks (diagnosed via their $H - K$ color excess) were mostly 
slow rotators, and that stars without $H - K$ excess were fast rotators. 

Following these initial findings, many observational studies have searched 
for signatures of the disk-locking effect by seeking to detect 
differences between the characteristic
rotation rates of stars that possess and lack circumstellar disks, under the
assumption that star-disk interactions will force stars with disks to rotate
more slowly than those stars that lack disks
\citep[e.g.][]{Herbst:2002aa,Rebull:2004aa,Covey:2005aa,Cieza:2007aa}.
The portrait of a star's angular momentum evolution that has emerged from
these efforts suggests that star-disk interactions lock a star to a
slow rotation period (P$\sim$8~d) matched to the angular velocity
of the disk's inner edge; in this picture stars spin up to become fast
rotators (P $\sim$1--2~d) only once their disks have begun to dissipate.
This picture suggests that slowly rotating young stars should possess
disks with smaller inner holes than their faster rotating contemporaries,
whose disks have presumably evolved such that star-disk interactions are no
longer able to govern the star's rotation rate.  

While most of these observational studies have tested mainly for a
statistical correlation between a young star's rotation period and the
presence or absence of a circumstellar disk, a key, generic prediction of 
disk-locking theories is that disk-locked stars should possess circumstellar 
disks with inner truncation radii ($R_{trunc}$) very nearly coincident with their 
co-rotation radus ($R_{co}$), the location where a Keplerian orbit within the disk
possesses the same angular velocity as the star's surface.
A few studies have attempted detailed comparisons of $R_{trunc}$
vs.\ $R_{co}$ for samples of young stars where these quantities could be
measured or inferred \citep[see][and references therein]{Carr:2007aa}. 
For example, \citet{Najita:2003} spectroscopically measured $R_{trunc}$
for six stars in Taurus-Auriga, finding that on
average $R_{trunc} \approx 0.7 \times R_{co}$. In the context of 
magnetic star-disk interaction models, this result would suggest that 
these stars are in fact experiencing active {\it spin-up} torque from 
their disks, since the stars would then be coupled to disk material with
higher specific angular momentum than the stars'.
However, the sample sizes remain too small to draw robust 
conclusions. Perhaps most importantly, the
range of important parameters---especially stellar rotation 
period---remains to be fully probed by such analyses. Indeed, the
\citet{Najita:2003} sample includes only slowly rotating stars with
$P_{rot} =$ 5--12 d. Thus, the role of star-disk interaction for more
rapidly rotating stars remains an important question.

There are also open questions concerning the universality of the disk locking mechanism. 
\citet{Stassun:1999aa}, \citet{Herbst:2002aa}, and \citet{Cieza:2007aa}, for example, 
found that the lowest mass stars lack the bimodal rotation period distribution
traditionally interpreted as another signature of disk-locking. 
Additionally, \citet{Stassun:2001aa} investigated the structure of circumstellar
disks as a function of rotation period and questioned the idea of a simple dichotomy between
disked slow rotators and diskless rapid rotators.
From a theoretical standpoint, \citet{Matt:2010aa}, also 
found that models of star-disk interactions incorporating the
impact of open field lines were unable to reproduce the observed population
of slow rotators. They moreover found that, while
the bulk of the stars in their models possessed disks truncated
at $R_{co}$, that did not necessarily imply a zero-torque configuration
where the star is ``locked" at a constant rotation rate. 

In this study, we analyze the properties of 33 stars in IC~348 to test
the agreement between their rotation periods and the Keplerian orbital
periods at the inner edges of their circumstellar disks.  In particular,
we seek to test two implications of the commonly presented picture of the
disk locking phenomena: (1) How do $R_{trunc}$ and $R_{co}$
compare for a well populated ensemble of circumstellar disks, and (2)
do fast rotators possess disks with $R_{trunc} \gg R_{co}$,
as expected if the disks around fast rotators have evolved to
the point that star-disk interactions no longer govern their host star's
rotation rate?  In Sec.~\ref{methodology}, we describe the fundamentals
of the test, as well as the parameters needed (stellar properties, disk
properties inferred from photometry, and the model grid). We then apply the
test and report the findings in Sec.~\ref{results}. We discuss
the implications in Sec.~\ref{discussion} and conclude with a summary in
Sec.~\ref{summary}.

\section{Methods\label{methodology}}
We aim to conduct a quantitative test of a central prediction of disk
locking theories: Does a young star rotate with a period equal to
the Keplerian orbital period of its inner disk?  
To perform this test, we define two characteristic locations within the
circumstellar disk: the distance from the star to the disk's inner edge,
$R_{trunc}$, and $R_{co}$, the radius at which the Keplerian angular 
velocity in the disk equals the star's angular velocity \citep{Ghosh:1979ab, Shu:1994aa}.	
$R_{co}$ is calculated for each star as:
\begin{equation}\label{e1}
R_{co} = (G M_\star P_{rot}^2 / 4 \pi^2)^{1/3}
\end{equation} 
\noindent where $M_\star$ and $P_{rot}$ are the star's mass and rotation
period, respectively.

To measure the truncation radius of each young star's circumstellar disk,
we analyze the amount of excess emission detected from each star at
near- and mid-infrared wavelengths, arising from warm dust in the inner
circumstellar disk.  Specifically, we compare mid-infrared photometry from
the Spitzer Space Telescope, as well as ground-based optical and
near-infrared observations, to synthetic spectral energy distributions
(SEDs) computed from a grid of 200,000 Monte-Carlo models covering a
wide range of parameter space.  We conduct this analysis on young stars
in IC 348, a nearby, young cluster in the Perseus star forming region.
This compact, optically visible region is amenable to photometric surveys
at optical wavelengths, enabling efficient measurements of stellar rotation
via star-spot modulation of stellar light curves, and the construction of
SEDs sampling the short wavelengths dominated by the stellar photosphere,
as well as the longer wavelengths dominated by the circumstellar disk.

\subsection{Photometry from the Literature\label{sedphot}}
In this study,
we make use of the SED measurements compiled by \citet{Lada:2006aa}
(L06) in their study of circumstellar disks in IC~348.	L06 combined
ground-based broadband
RIJHK measurements with mid-IR photometry from the Infrared
Array Camera (IRAC) and Multiband Imaging Photometer (MIPS) on the Spitzer
Space Telescope to produce SEDs for $\sim$300 stars previously
identified as cluster members by \citet{Luhman:2003aa}.  The L06 SEDs
span 0.5--24 $\mu$m, providing good sensitivity to emission from the
stellar photosphere as well as the inner circumstellar disk.  L06 reported
the broadband photometry in magnitudes; we converted these into flux units
using standard passband zeropoints 
\citep{Cousins:1976aa, Cohen:2003aa,Fazio:2004aa},
which are summarized in Table~\ref{calibrationtable}.

\subsection{Stellar Properties\label{stellarproperties}}
\begin{figure}
\figurenum{1}
\includegraphics{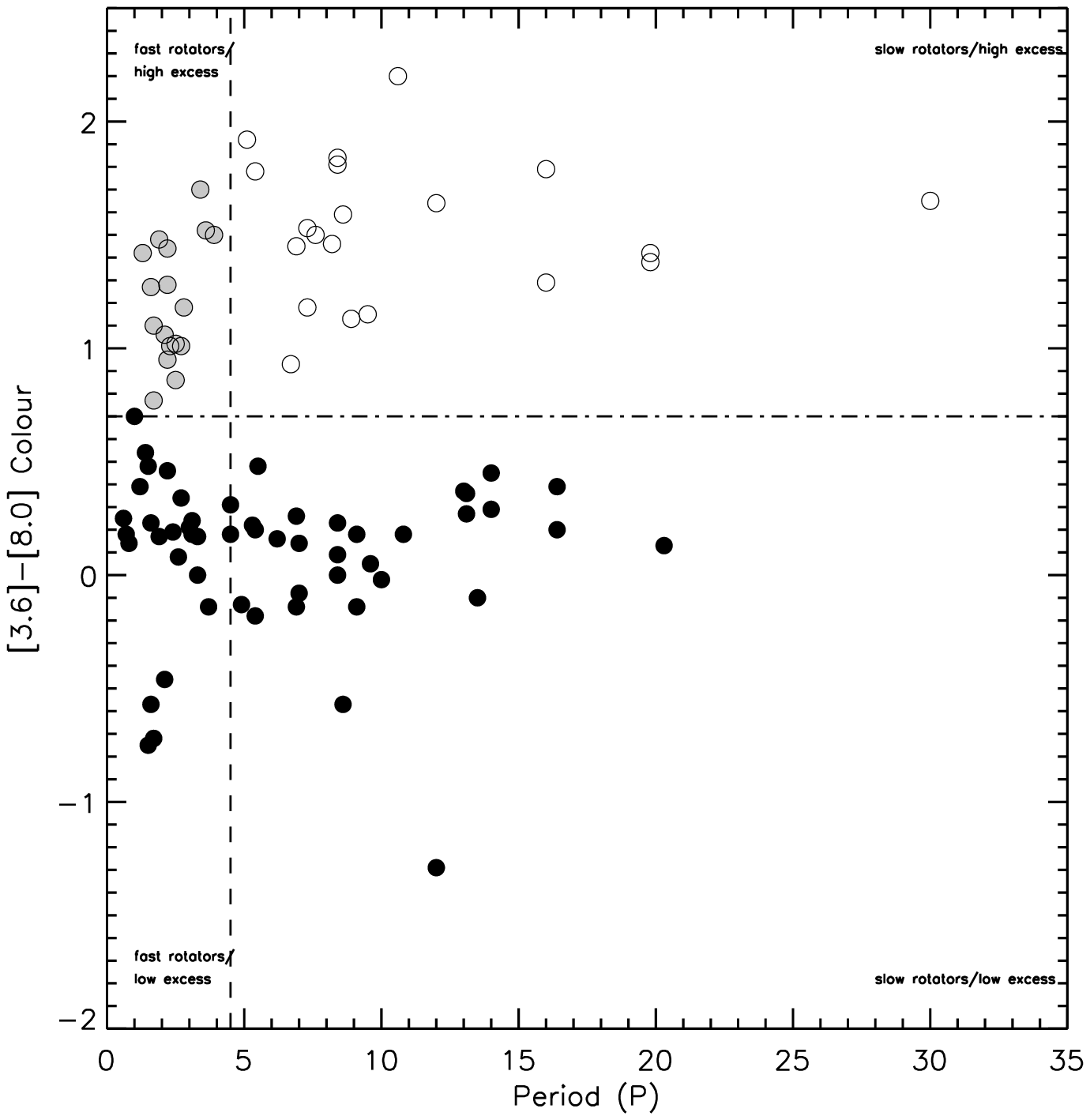}
\caption{IRAC [3.6]-[8.0] color vs. rotation period for IC 348 members.  The horizontal dashed line shows the IRAC color cut ([3.6]-[8.0] $>$ 0.7) used to identify IC 348 members with substantial circumstellar disks.  Periods are taken from the \cite{Lada:2006aa, Cieza:2006aa, Luhman:2003aa} catalogues: the vertical dashed line indicates a period of 4.5 days, used to separate the sample of stars with disks into subsets of fast rotators ($P_{rot} <$ 4.5 days; light grey filled circles) and slow rotators ($P_{rot} >$ 4.5 days; empty circles). See electronic edition of the journal for a coloured version of this figure.}
\label{colorvsperiod}
\end{figure}

Calculating $R_{co}$ for each star in our sample requires measurements of $P_{rot}$, $M_\star$, and $R_\star$
(see Eq. \ref{e1} above).
To infer each star's mass and
radius, we adopt the $T_{eff}$, $L_{bol}$ and $A_V$ values determined for
these stars by \citet{Luhman:2003aa}. Stellar masses were inferred for
each star by comparing the measured $T_{eff}$s and $L_{bol}$s to pre-main
sequence evolutionary models calculated by \cite{Dantona:1997aa}(DM97). We
calculate stellar radii using the fundamental Stephan-Boltzmann law,
which relates the star's luminosity ($L_\star$) to a given radius 
($R_\star$) and $T_{eff}$:
$L_\star = 4 \pi R_\star^2 \sigma T_{eff}^4$,
where $\sigma$ is the Stefan-Boltzmann constant. The full set of adopted and
inferred parameters for each star are presented in Table~\ref{derivedtable}.
Typical errors in $M_\star$ and $R_\star$ are derived from errors in
$T_{eff}$ and $L_\star$, with typical $T_{eff}$ errors of half a spectral
subtype ($\pm$82.5~K for M-type, $\pm$140~K for K-type), 
and typical errors of $\approx$0.3 in $\log L_\star$
\citep[e.g.][]{Hartmann:2001aa}.
We adopt rotation periods from the
catalog presented by \citet{Cieza:2006aa}, who measured $P_{\rm rot}$
due to starspot modulation of each star's light curve, from
their multi-epoch $I_C$ photometry of IC~348.  To identify stars with 
circumstellar disks, we applied a [3.6]$-$[8.0] $>$ 0.7 color cut 
as adopted by \citet{Cieza:2006aa} to the 
stars in the L06 catalog (see Fig.~\ref{colorvsperiod}). 
For the low-mass stars under consideration here, this color cut ensures 
that dusty disk material is present within $\sim$1~AU of 
our sample stars. 

This leaves 33 stars with the requisite data for our study.
For 27 of these we are able to determine the disk properties via
SED fitting (see below). These are summarized in Table~\ref{derivedtable}.

\subsection{SED Models\label{modelgrid}}
We make use of a pre-computed
grid of models, generated by \citet{Robitaille:2006aa} (henceforth R06),
to compare with photometry for stars in this study.  This grid builds upon
previous work done by \citet{Whitney:2003ab,Whitney:2003aa} 
(hereafter W03a, W03b),
by calculating the temperature structure of circumstellar disks of young
stellar objects (YSOs). W03a generated two-dimensional radiative transfer
models of Class~I YSOs, while W03b presented model SEDs, polarizations,
and images for an evolutionary sequence of YSOs from Class~0 to Class~III.
The W03a/W03b code uses Monte Carlo radiative equilibrium to generate
model SEDs for YSOs, following the trajectory of photon packets emitted
from a central stellar source into a disk and modeling their absorption,
re-emission, and/or scattering \citep{Bjorkman:2001aa}. This method yields
a model temperature structure specific to the parameters describing the
star (e.g., $T_{eff}$ and $R_*$) and its disk (e.g., $R_{trunc}$, $M_\star$, 
and scale-height).

The R06 model grid consists of SED models calculated using the W03a
algorithm, and covering a wide range of masses (from 0.1 to 50
$M_\sun$) and stages of YSO evolution.	R06 characterized
each model using 16 stellar, disk, and envelope parameters; the
most pertinent to this study include stellar mass, temperature,
and radius, as well as disk mass and $R_{trunc}$. The grid
consists of 200,000 SEDs computed at ten different angles (ranging
from near face-on at $18^\circ$, to near edge-on at $87^\circ$),
resulting in a comprehensive set of SEDs suitable to comparing with
actual YSO photometry. By comparing these synthetic SEDs with the
observed SEDs we have assembled for our sample, we can infer the
physical properties of each star's disk.

We used the R06 model grid to identify those models which reproduce
each IC 348 member's observed SED.  We limit each star's acceptable
fits, however, to those models with distances ($315 \pm 30$ pc) 
comparable to those
measured for most IC 348 members. The initial matches were further
screened on the basis of goodness of fit
with the observed SED, agreement with the $T_{eff}$ value reported
in the literature for each star, and the implied A$_V$ and disk mass:

\begin{itemize}
  \item{\textbf{T$_{eff}$ filter} --- We retain only those models with 
  $T_{eff}$ within $\pm$500~K of the value reported in the literature 
  from previous spectroscopy-based determinations.};

  \item{\textbf{$\chi^2$ filter} --- A $\chi^2$ metric is applied to ensure
  goodness-of-fit, such that the models selected possessed $\chi^2 \leq
  \chi^2_{best} + 9.21$. This value corresponds to a 99\% confidence level
  for two model parameters of interest in our SED fitting \citep{Press:1992aa},
  $R_{trunc}$ and $M_{disk}$ (see below).
  }

  \item{\textbf{$A_V$ filter} --- Previous work generally found 
  $A_V$ for IC~348 members to be modest, rarely larger than $\sim$5 mag 
  (e.g.\ L06). Therefore, our SED fitting results with A$_V\geq 10$ were
  eliminated to remove models with excessive combined interstellar and
  stellar extinction. In most cases, models with artificially large
  $A_V$ were already eliminated by the $T_{eff}$ filter above.
  }

  \item{\textbf{M$_{disk}$ filter} --- Models lacking sufficiently massive
  disks (M$_{disk}$ $\leq 10^{-4} M_{\sun}$) are also removed. Such low-mass
  disks are rarely seen in sub-mm surveys \citep{Andrews:2005aa}, and are 
  unlikely to be capable of sustaining significant star-disk interaction.}

\end{itemize}
	
\begin{figure}
\figurenum{2}
\centering
\includegraphics[angle=90, scale=0.75]{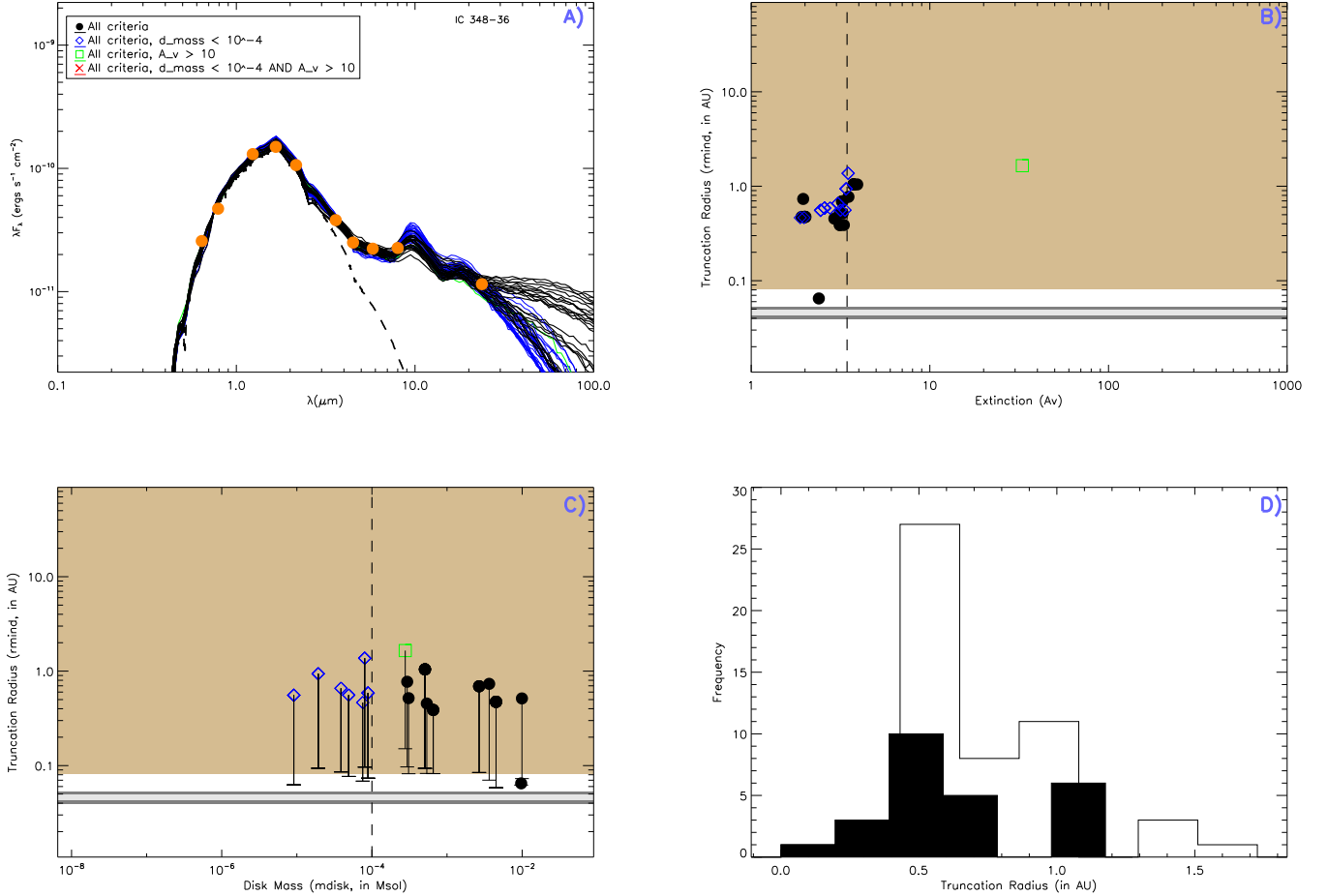}
\caption{SED fitting results for IC 348 36, an example of a star whose model fit predicts $R_{trunc} \gg R_{co}$, and $R_{trunc} \gg R_{sub}$.  \textbf{Panel A)} Photometric SED (detections shown as yellow filled circles; upper limits as yellow arrows) compared to an artificially reddened Phoenix stellar atmosphere with the same T$_{eff}$ (dashed line) and SED fits from the R06 model grid: black lines show R06 model SEDs meeting all criteria outlined in \S~\ref{modelgrid}, with colored lines showing models that fail one of those criteria (see legend in panel).  \textbf{Panels B and C)} Location of model fits in R$_{trunc}$ vs. M$_{disk}$ or A$_V$ parameter space.  Black points indicate models meeting all criteria in \S~\ref{modelgrid}.  Models failing the limits on M$_{disk}$ \& A$_v$ are shown as blue diamonds and green squares, respectively, with models failing both criteria shown as red crosses.  Lower ``error" bars indicate the distance between R$_{trunc}$ and R$_{sub}$ (at each model's T$_{eff}$). Vertical dashed lines in panels B \& C show the M$_{disk}$ limit and the A$_V$ value reported for this star in the literature, respectively. The domain where R $>$ R$_{sub}$ is indicated with a mocha background; light grey bars show the range of possible R$_{co}$s assuming a 50\% uncertainty in M$_*$, with dark grey bands indicating the range of possible R$_{co}$s assuming a conservative 100\% uncertainty in M$_*$. \textbf{Panel D)} Distribution of R$_{trunc}$ values for all R06 models satisfying the basic $\chi^2$ criteria (open histogram), and for models meeting all criteria (filled histogram). Coloured plots for all sources are available in the electronic edition of the journal.}

\label{ic348-36}
\end{figure}

\begin{figure}
\figurenum{3}
\centering
\includegraphics[angle=90, scale=0.75]{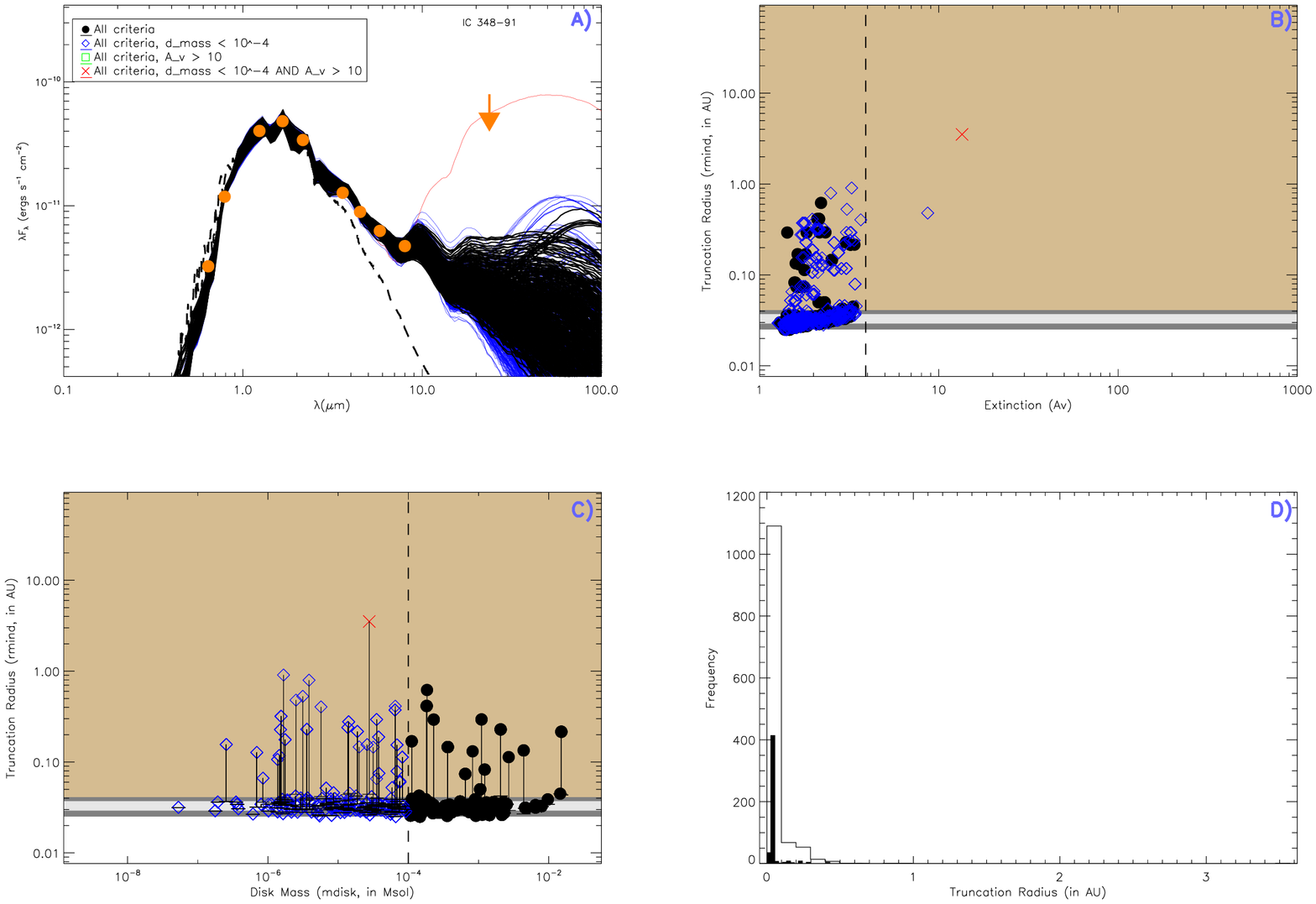}
\caption{IC 348 91. This sample SED and accompanying panels represent a star for which the model grid predicts $R_{trunc} = R_{co}$. Refer to Figure \ref{ic348-36} for further explanation of each of the above panels. Coloured plots for all sources are available in the electronic edition of the journal.}
\label{ic348-91}
\end{figure}

\begin{figure}
\figurenum{4}
\centering
\includegraphics[angle=90, scale=0.75]{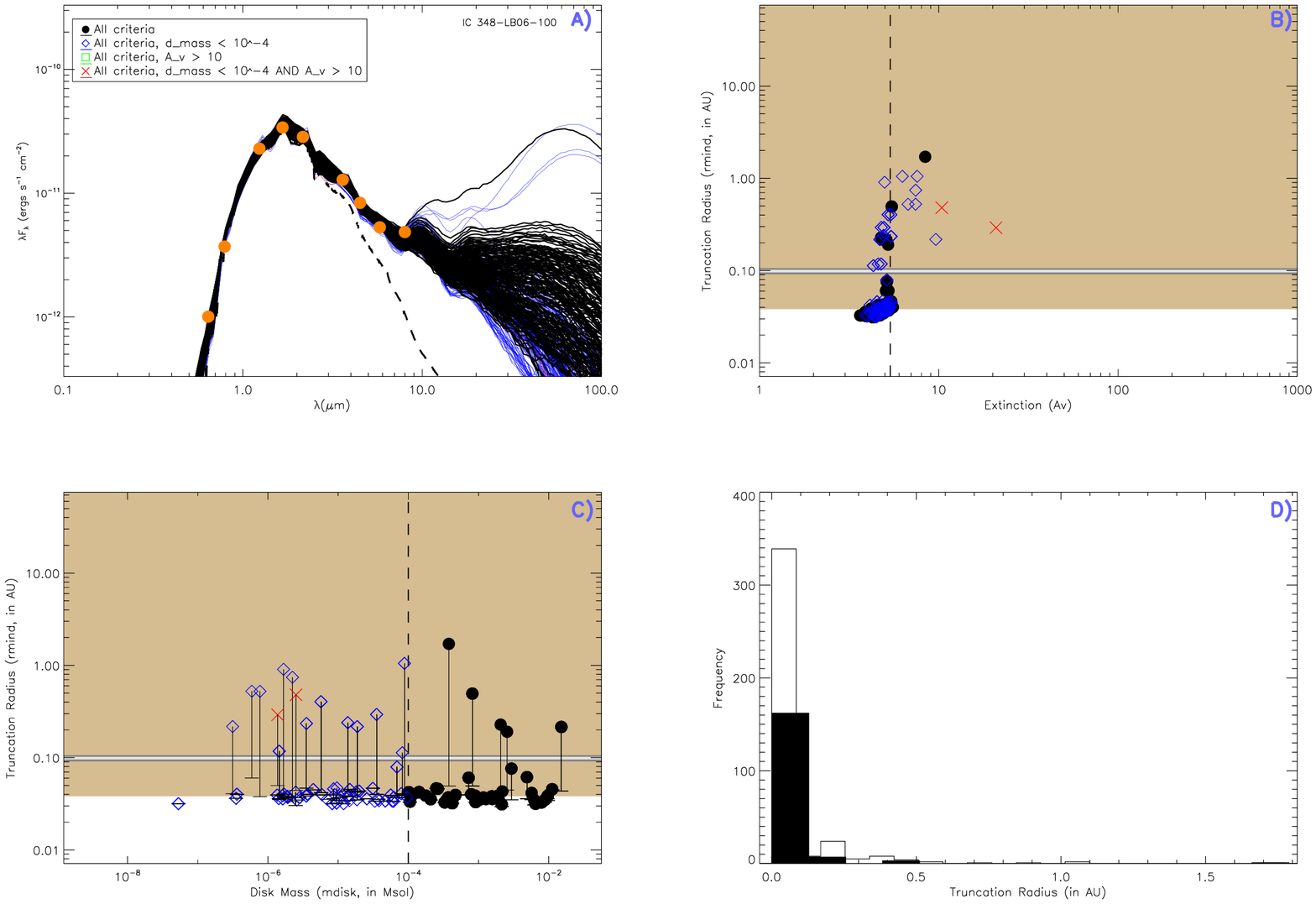}
\caption{IC 348 LB06-100. This sample SED and accompanying panels represent a star for which the model grid predicts $R_{trunc} = R_{sub}$. Refer to Figure \ref{ic348-36} for further explanation of each of the above panels. Coloured plots for all sources are available in the electronic edition of the journal.}
\label{ic348-LB06100}
\end{figure}

Of the 33 stars initially in our sample, six could not be matched to models
in the R06 grid which satisfy all of the above criteria (star IDs 6, 21, 41,
61, 140, and 182; refer to Table~\ref{discardedtable}). 
In most cases this was because the best-fit SED models required 
very low disk masses ($M_{disk} < 10^{-4} M_\sun$).
We therefore exclude these stars from our subsequent analysis. 
The results of the model vs.\ observed SED comparisons for the remaining
27 stars are discussed in \S~\ref{results}, with summaries of the
SED fitting results for three representative stars shown in 
Figs.~\ref{ic348-36}--\ref{ic348-LB06100}.

\section{Results}\label{results}

\subsection{Inferred Disk Truncation, Co-rotation, and Dust Sublimation Radii\label{radii}}

We have inferred $R_{trunc}$ for each star
by computing the mean $R_{trunc}$ of the full suite of R06
models which acceptably reproduce that star's SED and meet each of the
criteria outlined above. To provide context for these mean $R_{\rm trunc}$ values, 
we calculated the ratio between each star's $R_{\rm trunc}$ and its co-rotation 
and dust-sublimation radii ($R_{co}$ and $R_{sub}$, respectively).
The first of these ratios, $R_{\rm trunc}$/$R_{\rm co}$, 
is of course the principal quantity that we seek to test, as
$R_{trunc} / R_{co} \approx 1$ is
predicted by most disk-locking theories (see \S\ref{intro}).

We also compute
$R_{\rm trunc}$/$R_{\rm sub}$, which indicates if the $R_{\rm trunc}$
value returned by the SED model fits corresponds to the true inner edge of
the circumstellar disk.  
Magnetic star-disk interaction requires the stellar magnetic field lines 
to connect to ionized gas in the circumstellar disk.  
It is therefore important to note that the observed SEDs used here,
based on broadband fluxes, strictly speaking trace only the spatial extent 
of a disk's dust. However, dust that is sufficiently close to the stellar 
surface is expected to be destroyed via sublimation. This effect is
included in the R06 SED model grid; if a disk would otherwise extend
inward of $R_{sub}$, the disk is forced to have $R_{trunc}=R_{sub}$.
$R_{sub}$ is given by
\citep{Tuthill:2001aa, DAlessio:2004aa, Whitney:2003ab, Whitney:2003aa}:
$R_{sub} = R_\star \times (T_{sub} / T_{eff})^{-2.1}$
where $T_{sub}$ is the temperature at which dust is destroyed 
by photoevaporation (the R06 grid assumes $T_{sub} =$ 1600~K). 

In cases for which we find $R_{trunc} = R_{sub}$, 
we assume that the dust has been truncated by sublimation, 
a process which would not remove the gas
\citep[e.g.][]{Najita:2003,Eisner:2005}. 
Therefore, in these cases we assume that the gas in the disk in fact 
extends closer to the star than inferred from the observed SED; the 
inferred $R_{trunc}$ in these cases is therefore an upper limit.
Conversely, in cases for which we find $R_{trunc} > R_{sub}$,
some other process may be responsible for clearing out the inner 
portion of the disk, and therefore we
assume that the inner gas is cleared out as well 
\citep[e.g.][]{Isella:2009}.

\begin{figure}
\figurenum{5}
\centering
\includegraphics[scale=0.75]{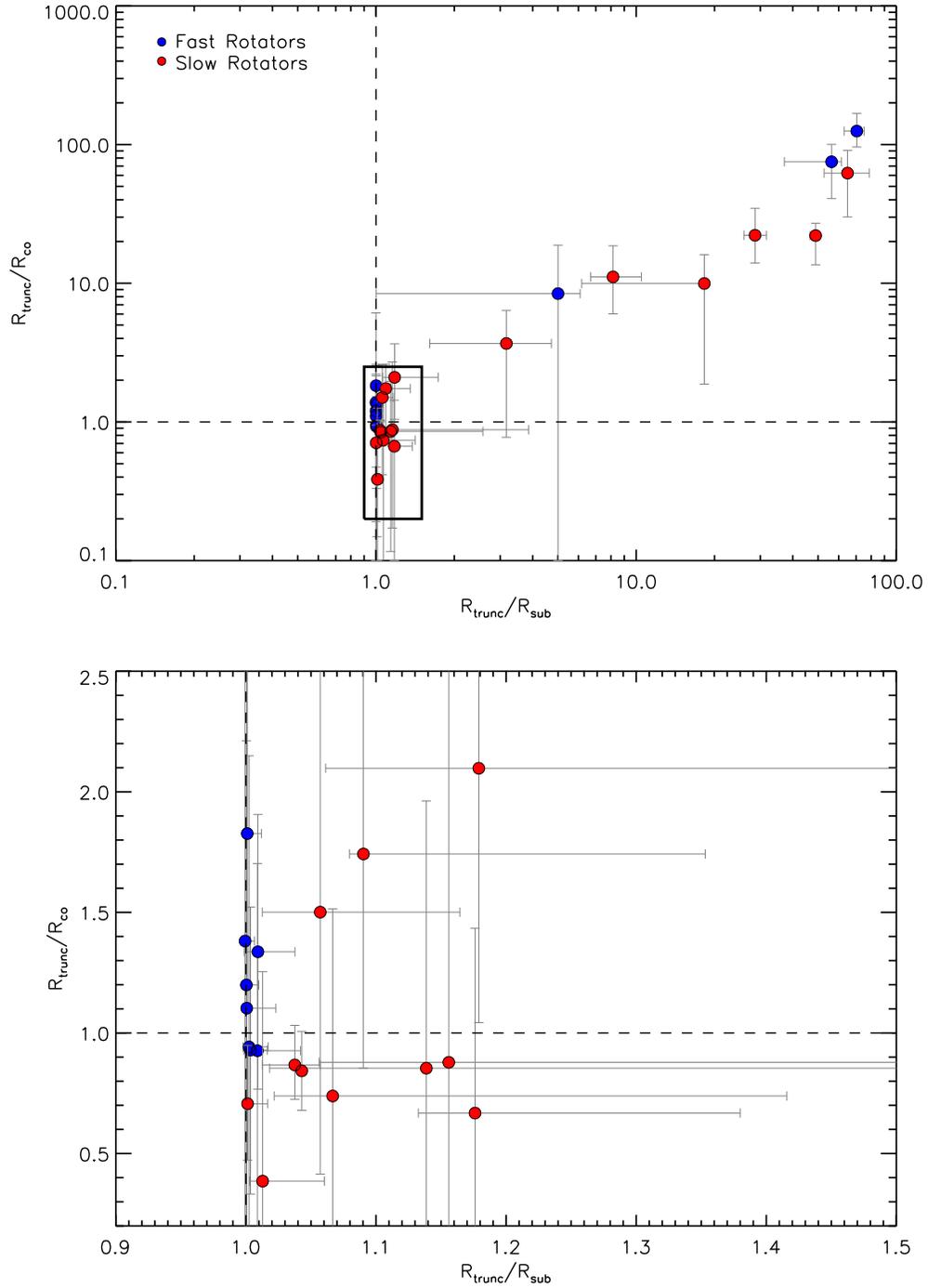}
\caption{Truncation with Corotation and sublimation radii ratio plot. The plot shows the ratios of the corotation and sublimation radii in relation to the mean truncation radii for all acceptable models in the R06 grid for the entire sample (top panel). The targets within the solid box are stars with potential disks (top panel, zoomed in bottom panel), while those outside are stars that are effectively diskless. Errors in the truncation/sublimation radius ratio (x-axis) are represented as interquartile range errors, with 25\% on the left and 75\% on the right of each data point. Truncation/corotation radius ratio errors are represented as 1 $\sigma$. See electronic edition of the journal for a coloured version of this figure.}
\label{ratioplot}
\end{figure}

Fig.~\ref{ratioplot} shows the $R_{\rm trunc}/R_{\rm co}$ and $R_{\rm
trunc}/R_{\rm sub}$ ratios for our entire sample.  We immediately
identify two distinct populations of stars: One group with 
$R_{trunc} \gg R_{co}$ and $R_{trunc} \gg R_{sub}$
(32\% of the final sample), and a second group with
$R_{trunc} \approx R_{co}$ (68\% of the sample). 

Since $R_{\rm co}$ is generally the location of ``action" in most
magnetic star-disk interaction models \citep{Shu:1994aa, Mohanty:2008aa,
Matt:2010aa}, the first group represents stars for which a magnetic 
star-disk interaction is most likely not important.  With physically very 
large $R_{trunc}$, far beyond $R_{co}$, these stars evidently harbor 
disks that have evolved significantly and no longer present 
substantial disk material within reach of the stellar magnetosphere;
we refer to these stars as ``effectively diskless".
In contrast, the latter group represents stars with substantial disk
material situated at or very near to the location of potential star-disk 
interaction. While the precise location of the inner-disk edge relative
to $R_{co}$ requires a detailed examination of possible dust
sublimation effects (which we do below), as discussed above the effect of 
such dust sublimation will be to imply a true $R_{trunc}$ that is even
closer to the star that what we have inferred, and for
which we might expect active interaction between the disk and the 
stellar magnetosphere to be even more likely. Therefore, 
we refer to this group of stars as ``potentially disk-locked".
We discuss the implications of these two groups in more detail below.

\subsection{Comparison of Potentially Disk-Locked and 
Effectively Diskless Stars}\label{hrdiagrams}
\begin{figure}
\figurenum{6}
\centering
\includegraphics{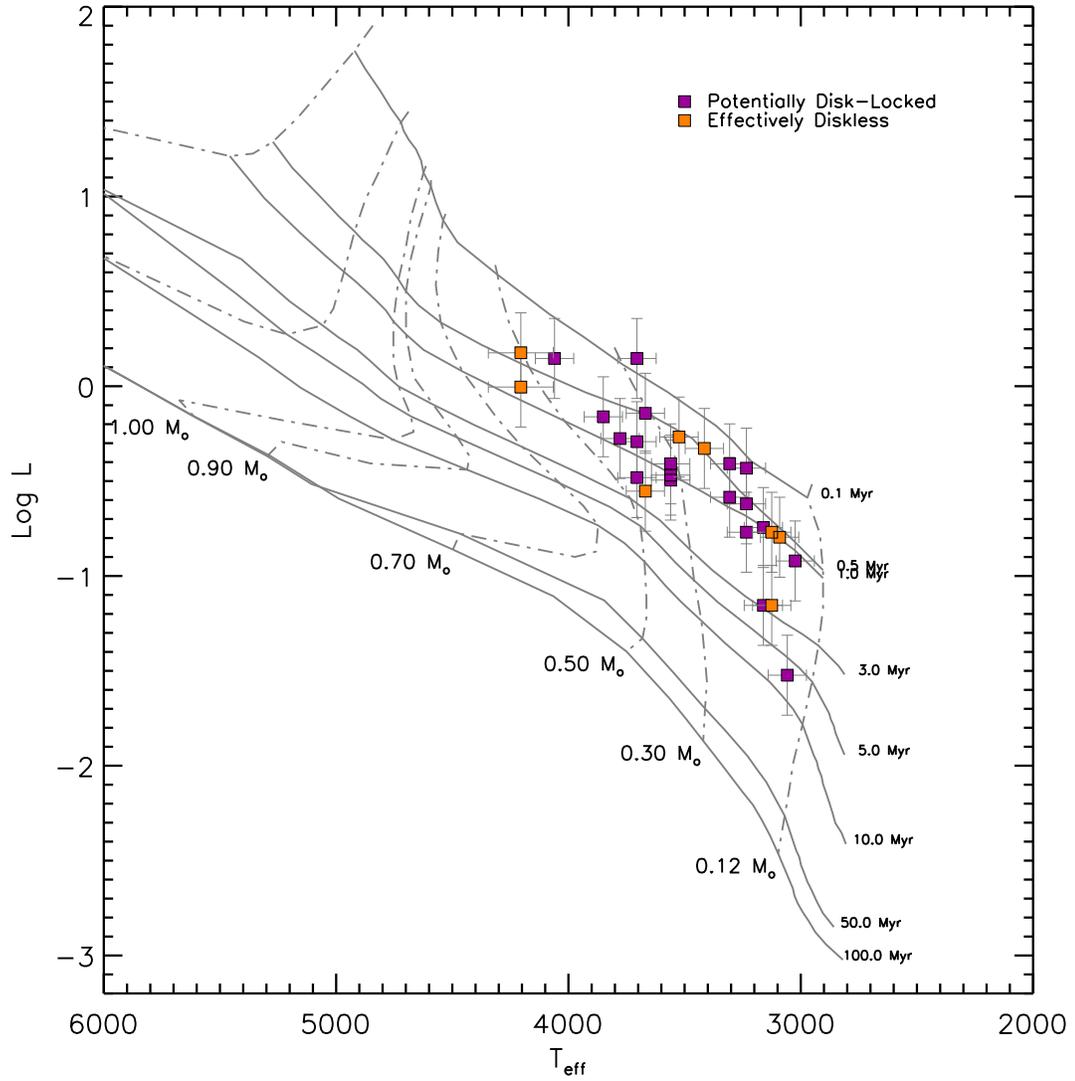}
\caption{HR Diagram with over plotted tracks and isochrones using DM97. All errors represented in both log L \& log T are 1 $\sigma$. See electronic edition of the journal for a coloured version of this figure.}
\label{ic348HRD}
\end{figure}

\begin{figure}
\figurenum{7}
\includegraphics{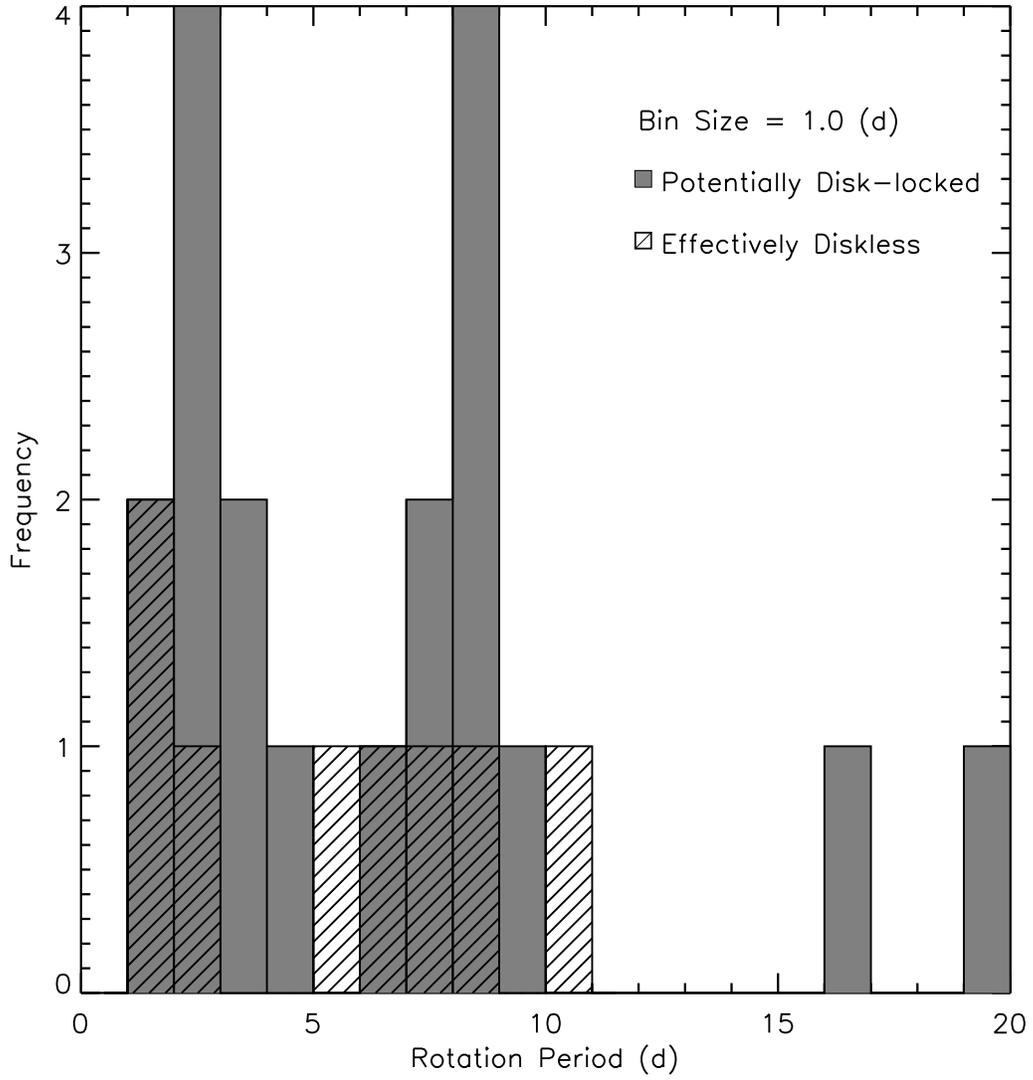}
\caption{Rotation period histogram of potentially disk-locked (gray) and effectively diskless stars (empty hatched). Though the potentially disk-locked stars are greater in number than the effectively diskless stars, this histogram supports the null hypothesis that both distributions are from the same parent distribution.}
\label{prothist}
\end{figure}

Figure \ref{ic348HRD} shows the location of these IC 348 members within the HR diagram,
with tracks and isochrones calculated by DM97 overlaid for comparison.
The IC 348 members analyzed here possess HR diagram locations consistent
with 1--2 Myr isochrones, with implied masses of 0.7 $M_{\sun}$ or below. 
There is no clear difference between the ages and masses of the members of 
the potentially disk-locked and effectively diskless groups:
this visual conclusion is supported by 1D and 2D two sided K-S tests, which indicate that the 
ages and masses of the stars in the two groups are consistent with shared
parent populations at 94\% and 74\% confidence levels (1D) and 52\% (2D). We also applied a two-sided K-S test on the rotation period distributions for the potentially disk-locked and effectively diskless groups (see Fig. \ref{prothist}). The K-S test in this case returns a probability of 0.74, and as such we cannot reject the null hypothesis that the two distributions are drawn from the same parent distribution.

\section{Discussion}\label{discussion}

Our SED modeling provides the first detailed investigation of how the structure
of circumstellar disks around IC 348 members does or does not influence the star's rotation rate. 
Specifically, these measurements provide leverage to address the two
questions motivating this study: 1) how do the inner truncation radii and
co-rotation radii compare for a well populated ensemble of circumstellar
disks, and 2) do fast rotators possess circumstellar disks with inner radii
larger than co-rotation, as expected if the disks around fast rotators have
evolved to the point that star-disk interactions no longer govern their
host star's rotation rate?  We address each of these questions in turn.

\subsection{Are circumstellar disks truncated at co-rotation?}

Traditionally, the condition $R_{\rm trunc} \approx R_{\rm co}$ has
been assumed as a fundamental requirement for a star to be in a fully 
disk-locked state, where a quasi-steady-state configuration of the star-disk 
interaction exerts a braking torque on the star that counter-balances the star's
tendency to spin up as it contracts, and thus maintains a roughly constant
stellar rotation period for the lifetime of the disk. For example, in the
context of the bimodal distribution of rotation periods reported by Herbst
and collaborators in the ONC \citep[e.g.][]{Herbst:2002aa},
the slow rotators have been interpreted as
those in an actively ``disk-locked" state while the rapid rotators have been
interpreted as ``disk-released," presumably due to the loss of their disks.
These interpretations have generally been made on the basis of near- and mid-IR colors
as functions of stellar rotation period, not on an assessment of the $R_{\rm
trunc} / R_{\rm co}$ condition on a star by star basis.  

Our detailed SED analysis reveals that the majority (70\%) of the stars 
in our sample have clear evidence for 
$R_{\rm trunc} \approx R_{\rm co}$ (Fig.~5). 
As these potentially disk-locked stars constitute the majority of our 
sample, it does appear that most stars in IC~348 are
consistent with the zeroth order prediction of theoretical models of
angular momentum transfer via star-disk interactions.

However, assessing whether the potentially disk-locked stars are 
experiencing a {\it braking} torque from any presumed star-disk interaction
is made more subtle by the complicating effects of dust sublimation 
on the determination of the disk's true $R_{trunc}$.
As shown in Fig.~5, while both the slow and rapid rotators in the 
potentially disk-locked group are similarly
distributed along the vertical axis (i.e., both the slow and
rapid rotators have disks consistent with $R_{\rm trunc} = R_{\rm co}$,
within error), their distributions along the horizontal axis are detectably
different. The rapid rotators are very strongly clustered
at precisely $R_{\rm trunc}/R_{\rm sub} = 1$
(scatter in $R_{\rm trunc}/R_{\rm sub}$ of less than 1\%), 
which is the hard minimum that any of
our SED models can attain because (by definition) dust is destroyed by
sublimation interior to this radius. In contrast, the slow rotators show
a larger spread of $\sim$15\% in $R_{\rm trunc}/R_{\rm sub}$. Moreover,
most of the slow rotators' disks are mildly inconsistent with $R_{\rm
trunc} = R_{\rm sub}$, requiring $R_{\rm trunc} > R_{\rm sub}$ with greater
than 1$\sigma$ confidence. Taken as a group, the rapid rotators 
show a high likelhood of possessing a mean 
$R_{\rm trunc}/R_{\rm sub}$ = 1, while the likelihood that the slow
rotators possess a mean $R_{\rm trunc}/R_{\rm sub} = 1$ is less than 0.1\%.

We interpret the uniform pile-up of fast rotators at $R_{\rm
trunc}/R_{\rm sub} = 1$ to mean that our SED modeling is not sensitive
to the true inner edge of the fast rotators' disks, which likely extend
inward of $R_{sub}$ and perhaps significantly inward of $R_{co}$ as well
(though we cannot verify the latter).
In contrast, the $R_{\rm trunc}/R_{\rm sub}$
ratios that we observe for the slow rotators are significantly larger than 1.
Therefore we regard the $R_{\rm trunc}$ for the slow rotators 
in general to correspond to true $R_{trunc}$ measurements,
and thus we can conclude with greater confidence that these stars
truly have disks consistent with $R_{trunc}/R_{co} = 1$. 

If the potentially disk-locked stars in our sample are in fact experiencing
angular momentum interactions with their disks, the above findings could
imply that the slow rotators are slow precisely because they satisfy the
$R_{trunc} / R_{co} = 1$ condition, while the fast rotators are fast
because they tend to possess disks with $R_{trunc}/R_{co} < 1$.
(To be clear, our results do not demonstrate that $R_{trunc}/R_{co}<1$ 
for the rapid rotators in the potentially disk-locked group, but 
such an interpretation would be consistent with our findings above.)
Recent theoretical work complicates such a straightforward interpretation,
however. \citet{Matt:2010aa} have demonstrated that the condition $R_{\rm
trunc} / R_{\rm co} \approx 1$ can in fact transpire for a very wide range 
of star/disk parameters and a wide range of star-disk torque configurations.
Indeed, in those calculations, relatively
small deviations from $R_{\rm trunc}/R_{\rm co}=1$ can result in large
differences in the magnitude and/or the sign of the torque experienced by
the star. For example, in the case of strong magnetic coupling to the disk
(e.g., $\beta = 0.01$), significant field twisting occurs for $R_{\rm
trunc}$ deviations of less than 1\% from $R_{\rm co}$.
The $R_{\rm trunc}/R_{\rm co}$ that we have determined for potentially
disk-locked stars are not sufficiently precise to make such distinctions. 

Nonetheless, the findings of \citet{Matt:2010aa} do still predict that
disks with smaller $R_{trunc}/R_{co}$ will in general result in more
positive stellar torques.  
Thus, if the potentially disk-locked stars in our sample are in fact
disk-locked, the implication is that the rapid rotators are likely
experiencing systematically more positive torques.  More generally, 
under the disk-locking hypothesis, our results are most consistent with 
an interpretation in which the stars are currently experiencing disk 
torques spanning a large range of magnitude, and perhaps in sign.

\subsection{Do fast rotators possess (dust) disks with larger inner holes?}

As commonly envisaged, magnetic star-disk interactions force the star to
rotate at the Keplerian velocity of the close-in inner disk.  Over time,
however, the disk's inner hole grows\footnote{While other models of 
disk evolution that do not predict a widening 
inner-disk hole do exist (e.g., homologous depletion; 
\citep{Currie:2009ab, Currie:2011aa}, 
`inside-out' disk evolution is the most commonly invoked
\citep[e.g.][]{Barsony:2005}.}
as the disk evolves and begins to
dissipate; eventually star-disk interactions are too weak to couple the
star to the anemic inner disk, and the star spins up as it 
completes its pre-main sequence contraction.
This picture would predict that slowly rotating stars will be locked to 
disks with $R_{trunc} \approx R_{co}$, while rapidly rotating stars
would be associated with disks with $R_{trunc} \gg R_{co}$
\citep{Edwards:1993aa, Edwards:1994aa, Rebull:2006aa, Cieza:2007aa}.
To test this picture,
we can compare the $P_{rot}$ distributions for the potentially 
disk-locked and effectively diskless stars in our sample.

Interestingly, we find 
slow and rapid rotators in roughly equal numbers among the potentially disk-locked stars. This suggests
that the common interpretation, wherein fast rotators result from stellar
spin-up following the cessation of disk-locking, may need to be modified
to allow for the existence of rapidly rotating stars with close-in disks. 
Indeed, both slow and rapid rotators amongst the potentially disk-locked stars in our sample
at least approximately satisfy the condition $R_{\rm trunc} \approx R_{\rm co}$.
As discussed above, the rapid rotators in the potentially disk-locked
group may in fact possess disks with $R_{trunc} < R_{co}$, however this
only strengthens the conclusion that rapid rotators do not possess
disks with systematically larger inner holes compared to the slow rotators.

The effectively diskless stars in our sample 
are very unlikely to be experiencing significant torques
from their disks. Interestingly, however, the rotation period distribution
of these stars is nonetheless very similar to that of the potentially disk-locked stars 
(see \S\ref{hrdiagrams}).
Presuming that the effectively diskless stars formerly possessed robust 
inner disks like their present-day potentially disk-locked counterparts, 
this implies that their disk properties have evolved significantly 
while their rotational properties have remained unchanged. 
If the currently effectively diskless stars are furthermore
presumed to have formerly 
been in a disk-locked state, then the timescale for transitioning from the 
disk-locked to the disk-released state must be shorter than the timescale 
on which the stars would spin up due to pre-main-sequence contraction. This conclusion
is consistent with previous estimates of disk evolution timescales, which find timescales of 
$\sim$0.1-1 Myr for disk evolution processes \citep{[], Currie:2011aa, Muzerolle:2010aa}, compared to 
the $\sim$3 Myr spin-up timescale predicted by the DM97 PMS evolutionary models for 0.4$M_{\sun}$ 
stars (the average mass of our sample) at 1 Myr. This
is also consistent with the statistics of our sample: 1/3 of our sample are "effectively diskless", 
implying a disk evolution timescale (assuming the nominal age of 1 Myr typically adopted for
IC 348) of $\sim$1/3 $\times$ 1 Myr = $\sim$0.3 Myr, in agreement with aforementioned 
timescale estimates.

\section{Summary and Conclusions}\label{summary}

We have analyzed the circumstellar disks around a sample of 33 stars in IC
348 with known rotation periods, in order to assess in detail 
whether the inner edge of each star's circumstellar disk ($R_{trunc}$)
is consistent with being at the co-rotation radius from the star
($R_{co}$), as predicted by disk-locking
theory. We compare stellar photometry from the Spitzer Space Telescope
to a grid of 200,000 pre-computed star$+$disk radiative transfer models,
and compare the implied $R_{trunc}$ of the best fitting SED models to
each star's calculated $R_{co}$. The principal findings of this study 
are as follows:

\begin{itemize}

\item 
We find two populations of stars: a ``potentially disk-locked group"
with inner-disk radii located at $R_{\rm trunc}/R_{\rm co} \approx 1$ (68\% of the sample), 
and an ``effectively diskless" group whose inner-disk radii are 
significantly larger, with $R_{\rm trunc}/R_{\rm co} \gg 1$ and thus beyond the reach of disk-locking (32\% of the sample).

\item
Both fast and slow rotators in the potentially disk-locked group 
possess dust disks with
$R_{trunc}/R_{co}$ values consistent with 1. This finding is contrary to
previous suggestions that slowly-rotating stars will possess close-in 
disks that facilitate strong star-disk interactions, while fast rotators
will possess more evolved disks with inner radii that are sufficiently large
so that they are no longer amenable to significant star-disk interactions.
The general agreement between the each star's $R_{trunc}$ and $R_{co}$
may be taken to suggest that star disk-interactions do
indeed play a role in these stars' angular momentum evolution.  However, 
under this hypothesis,
the lack of a clear distinction between the $R_{trunc}$ inferred for
the disks around fast and slow rotators would imply that star-disk
interactions influence the rotation rate of the fast rotators as well, not
only the slow rotators whose periods disk locking is most commonly invoked
to explain.

\item
Stars in both the potentially disk-locked and effectively diskless
groups, whose disks we interpret as being
in significantly different evolutionary states, possess statistically
identical rotation period distributions. 
If the potentially disk-locked stars are presumed to currently be in a
disk-locked state, and if the effectively diskless stars are presumed 
to previously have been in a similarly disk-locked state, 
this suggests that disks evolve more quickly than the stellar 
spin-up timescale resulting from pre-main-sequence contraction.

\end{itemize}

In summary, while our findings may be interpreted within the context of 
a presumed disk-locking mechanism, invoking the disk-locking hypothesis 
is not necessitated by the stars in our IC~348 study sample. 
We do not find obvious differences in the disk truncation radii of 
slow rotators vs.\ rapid rotators. This holds true both at the
level of whether close-in disk material is present at all, 
and in analyzing the precise location of the inner disk edge
relative to the co-rotation radius amongst the subset of stars with 
close-in disk material. 
These results may therefore imply that the disk-locking phenomenon is not 
operative in these stars. 
Alternatively, if disk-locking does operate, then our findings imply
that (a) its observational signature is more complex than the simple 
portrait of slowly rotating disk-locked stars and rapidly rotating 
non-disk-locked stars, and (b) the transition from
the disk-locked state to the disk-released state must occur more rapidly
than the stellar contraction timescale.

\acknowledgments
We acknowledge and thank Sean Matt for helpful and insightful
discussions of angular momentum transfer via star-disk interactions. This
research has made use of NASA's Astrophysics Data System Bibliographic
Services, the SIMBAD database, operated at CDS, Strasbourg, France, and
photometry acquired as part of the Two Micron All Sky Survey.  The Two
Micron All Sky Survey was a joint project of the University of Massachusetts
and the Infrared Processing and Analysis Center (California Institute
of Technology). The University of Massachusetts was responsible for the
overall management of the project, the observing facilities and the data
acquisition. The Infrared Processing and Analysis Center was responsible
for data processing, data distribution and data archiving.
TSL gratefully acknowledges the support provided for this work by NASA
through the NASA Graduate Student Research Program (GSRP).
KRC gratefully acknowledges support provided for this work by NASA through Hubble Fellowship grant HST-HF-51253.01-A
awarded by the STScI, which is operated by the AURA, Inc., for NASA, under contract NAS 5-26555, and by the Spitzer Space Telescope Fellowship Program, through a contract issued by the Jet Propulsion Laboratory, California Institute of Technology under a contract with NASA. 


\bibliography{ms}

\appendix

\clearpage

\begin{figure}
\figurenum{8}
\centering
\includegraphics[angle=90, scale=0.75]{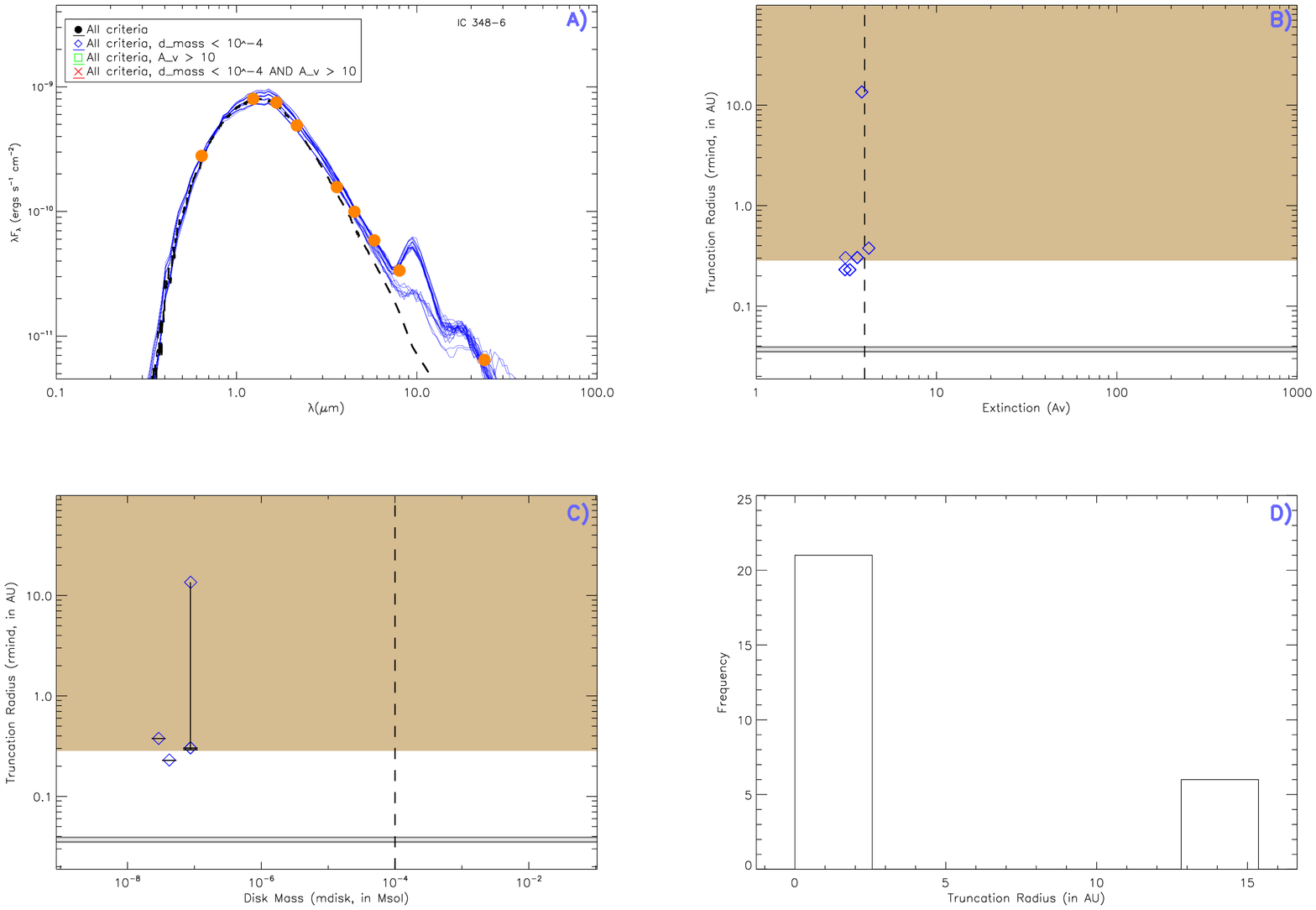}
\caption{IC 348 6. Refer to Figure \ref{ic348-36} for further explanation of each of the above panels. Figures 7 - 37 are available in the online version of the Journal.}
\label{ic348-6}
\end{figure}

\begin{figure}
\figurenum{9}
\centering
\includegraphics[angle=90, scale=0.75]{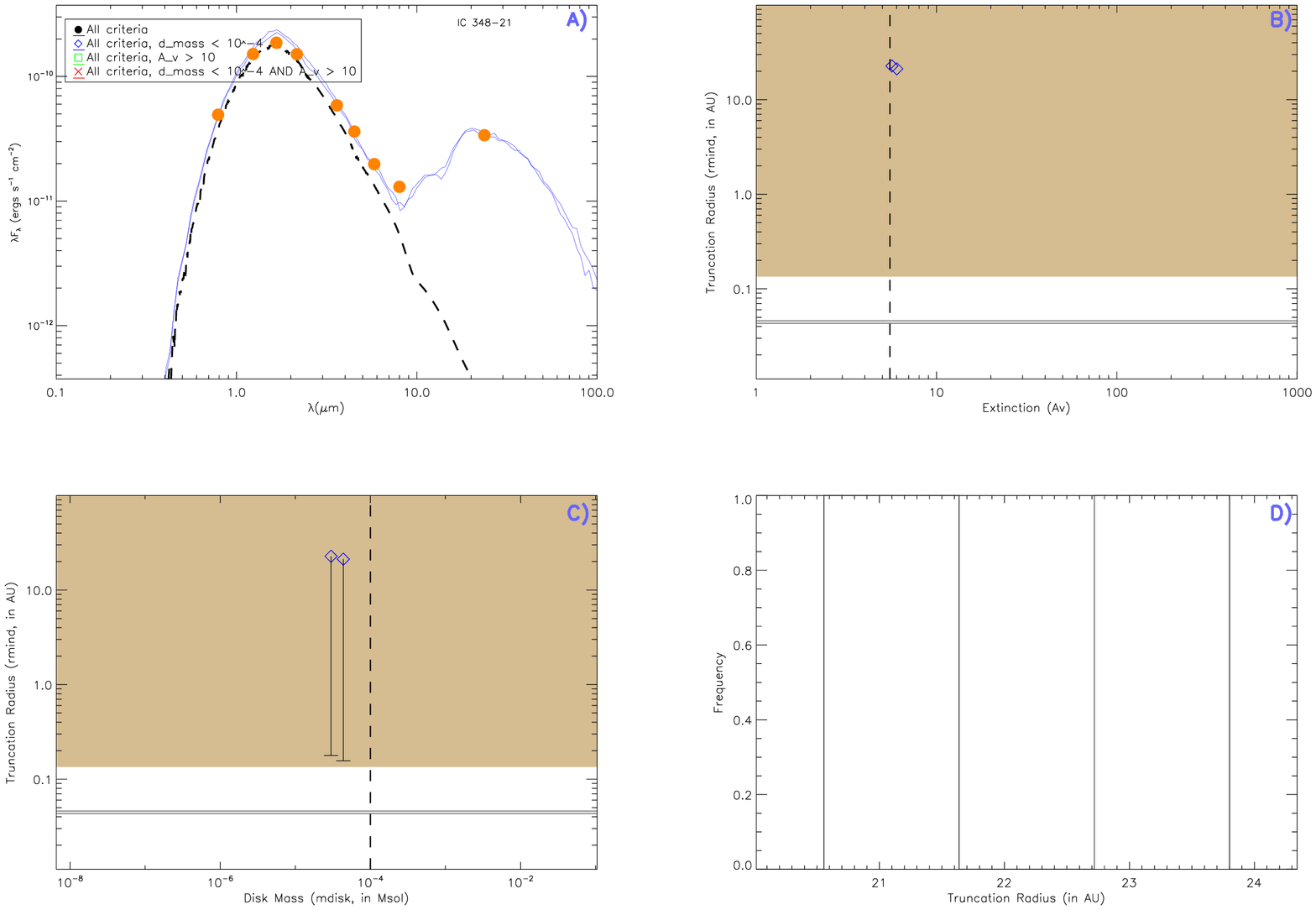}
\caption{IC 348 21. Refer to Figure \ref{ic348-36} for further explanation of each of the above panels. Figures 7 - 37 are available in the online version of the Journal.}
\label{ic348-21}
\end{figure}

\begin{figure}
\figurenum{10}
\centering
\includegraphics[angle=90, scale=0.75]{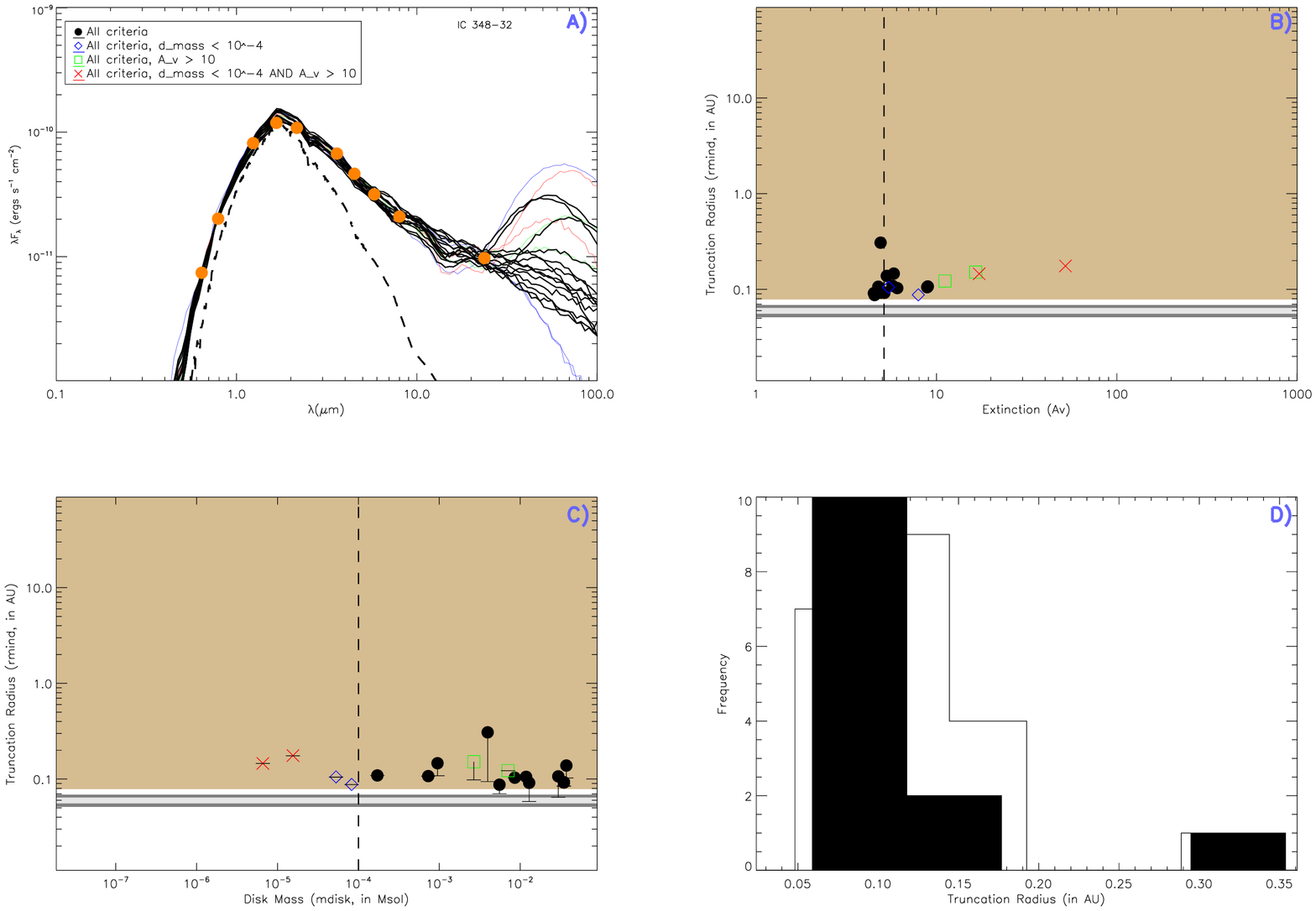}
\caption{IC 348 32. Refer to Figure \ref{ic348-36} for further explanation of each of the above panels. Figures 7 - 37 are available in the online version of the Journal.}
\label{ic348-32}
\end{figure}

\begin{figure}
\figurenum{11}
\centering
\includegraphics[angle=90, scale=0.75]{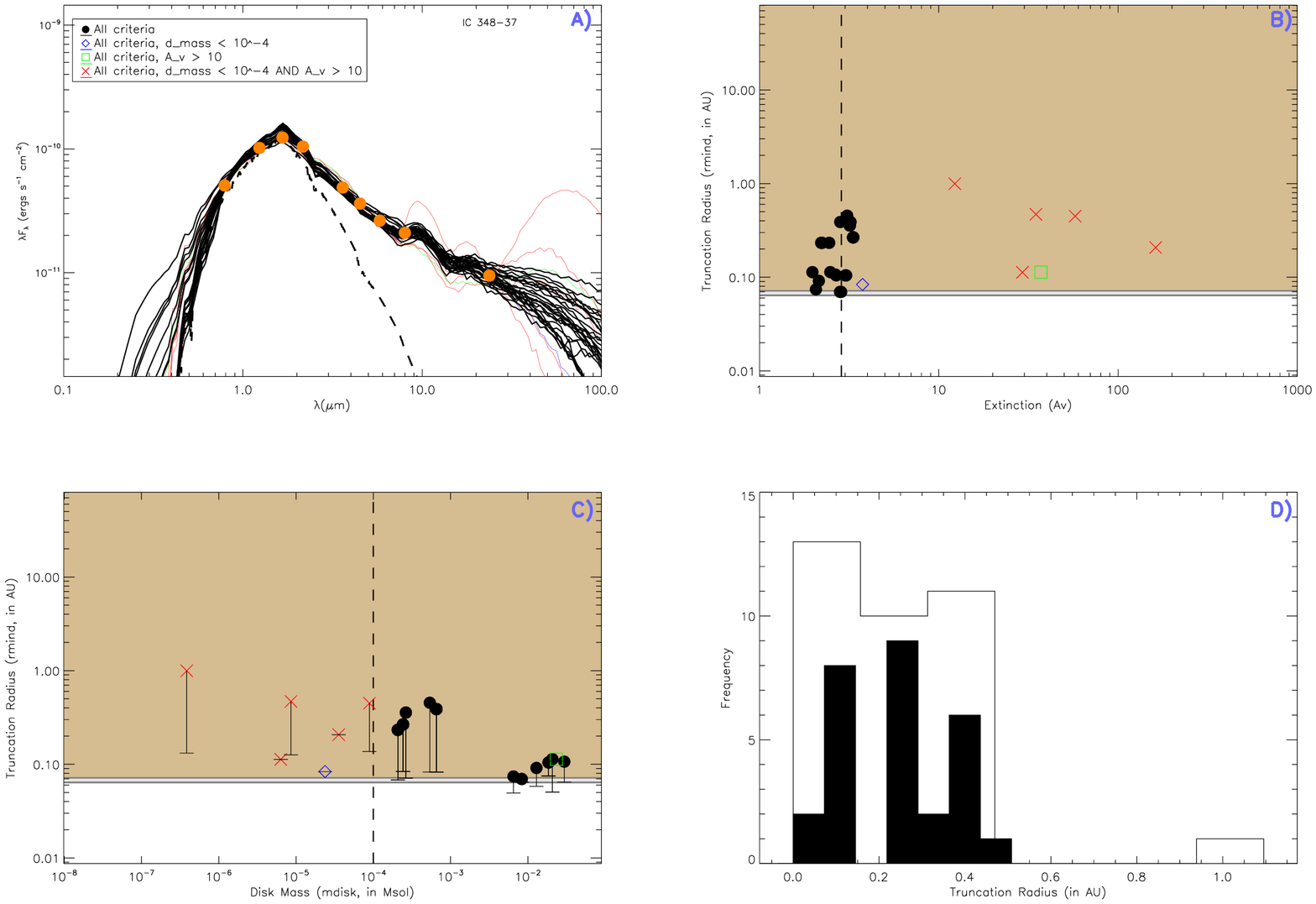}
\caption{IC 348 37. Refer to Figure \ref{ic348-36} for further explanation of each of the above panels. Figures 7 - 37 are available in the online version of the Journal.}
\label{ic348-37}
\end{figure}

\begin{figure}
\figurenum{12}
\centering
\includegraphics[angle=90, scale=0.75]{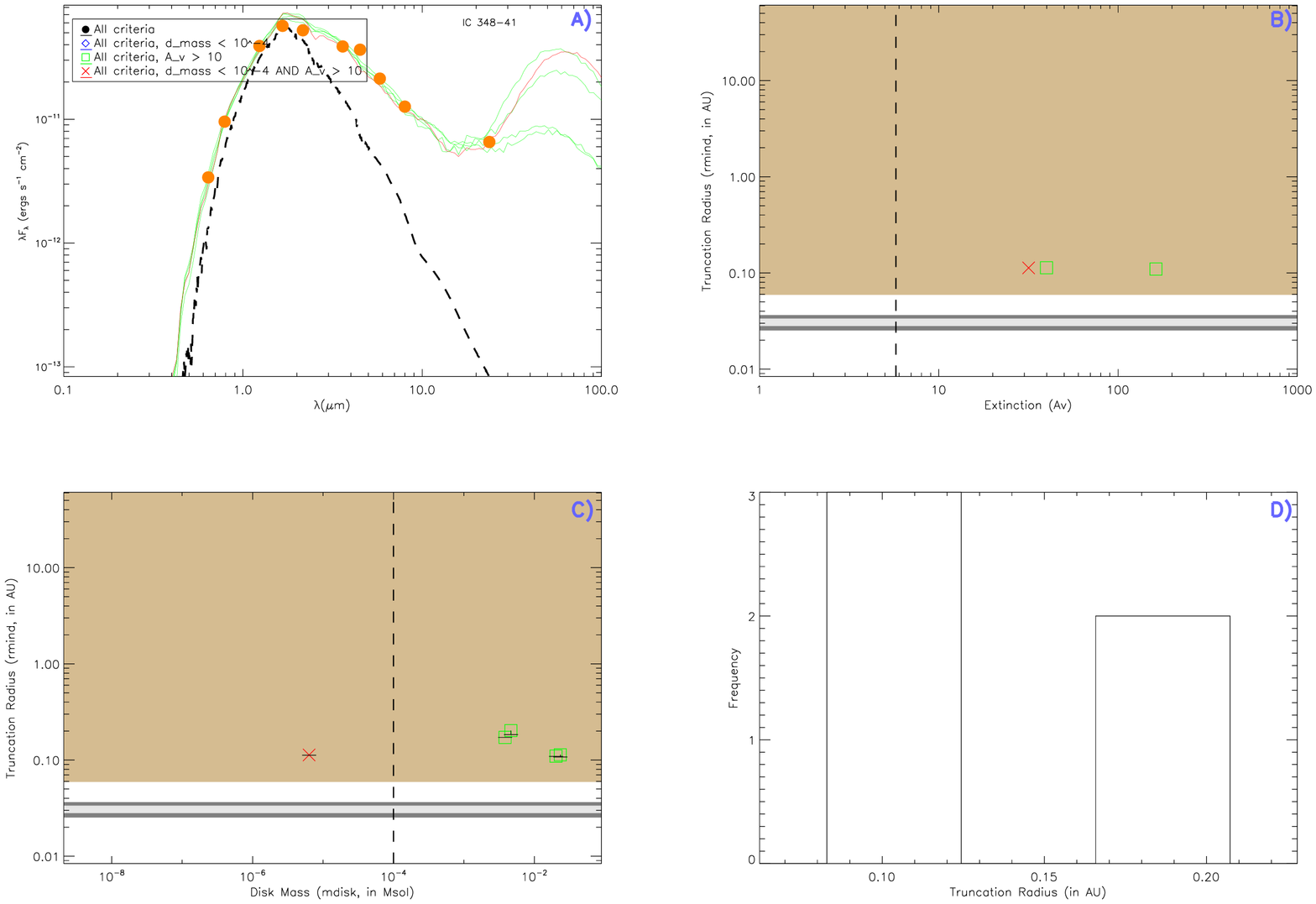}
\caption{IC 348 41. Refer to Figure \ref{ic348-36} for further explanation of each of the above panels. Figures 7 - 37 are available in the online version of the Journal.}
\label{ic348-41}
\end{figure}

\begin{figure}
\figurenum{13}
\centering
\includegraphics[angle=90, scale=0.75]{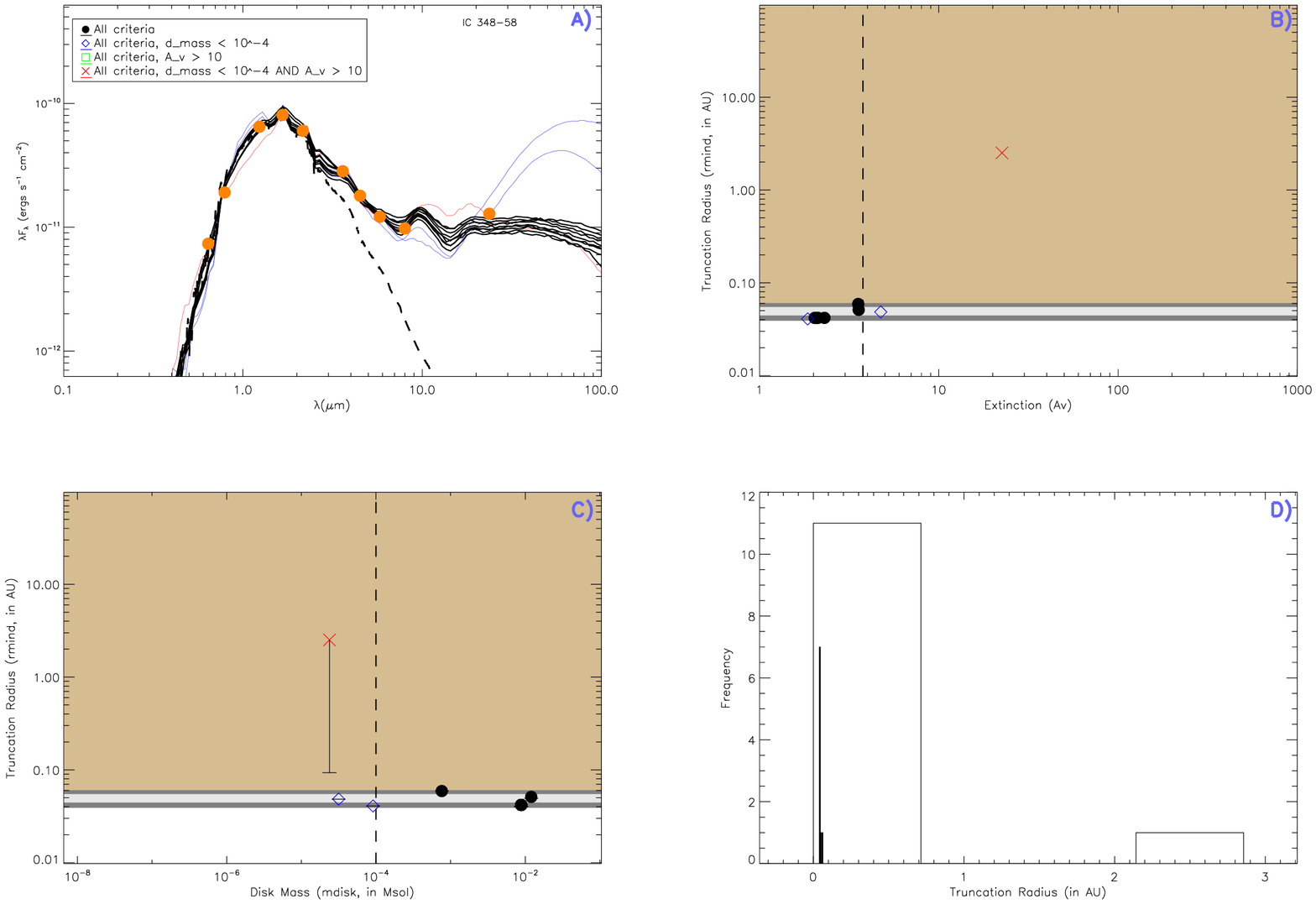}
\caption{IC 348 58. Refer to Figure \ref{ic348-36} for further explanation of each of the above panels. Figures 7 - 37 are available in the online version of the Journal.}
\label{ic348-58}
\end{figure}

\begin{figure}
\figurenum{14}
\centering
\includegraphics[angle=90, scale=0.75]{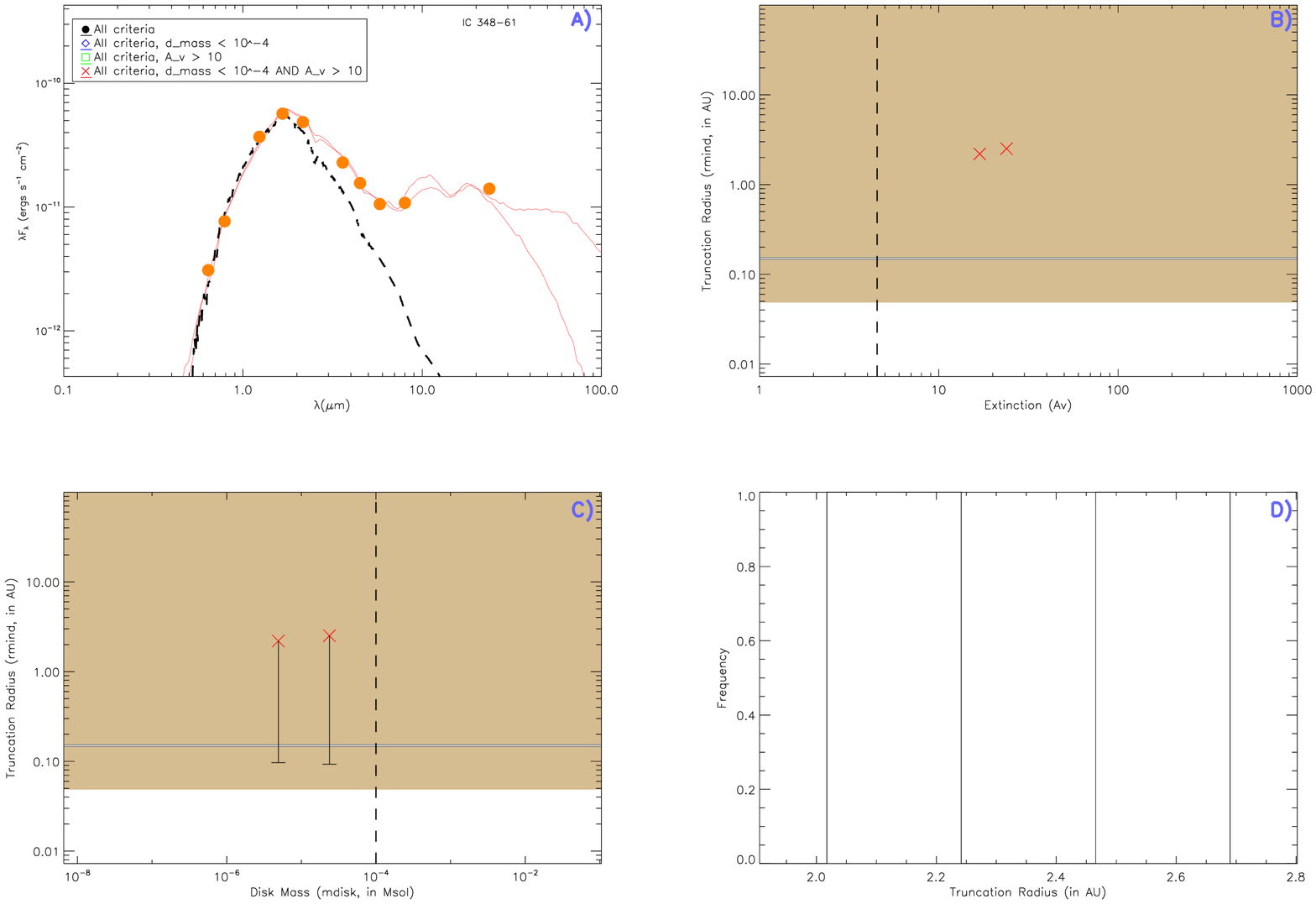}
\caption{IC 348 61. Refer to Figure \ref{ic348-36} for further explanation of each of the above panels. Figures 7 - 37 are available in the online version of the Journal.}
\label{ic348-61}
\end{figure}

\begin{figure}
\figurenum{15}
\centering
\includegraphics[angle=90, scale=0.75]{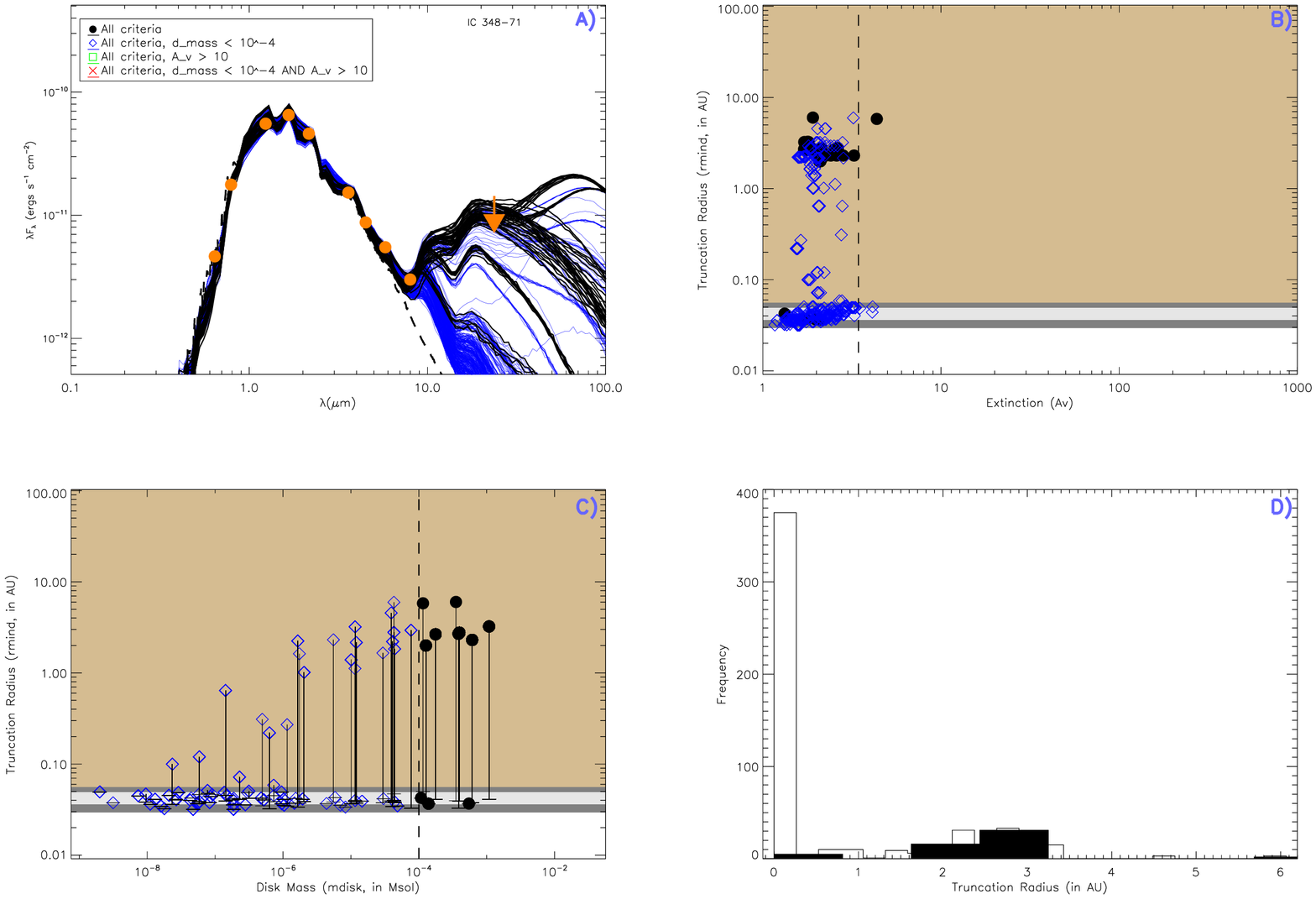}
\caption{IC 348 71. Refer to Figure \ref{ic348-36} for further explanation of each of the above panels. Figures 7 - 37 are available in the online version of the Journal.}
\label{ic348-71}
\end{figure}

\begin{figure}
\figurenum{16}
\centering
\includegraphics[angle=90, scale=0.75]{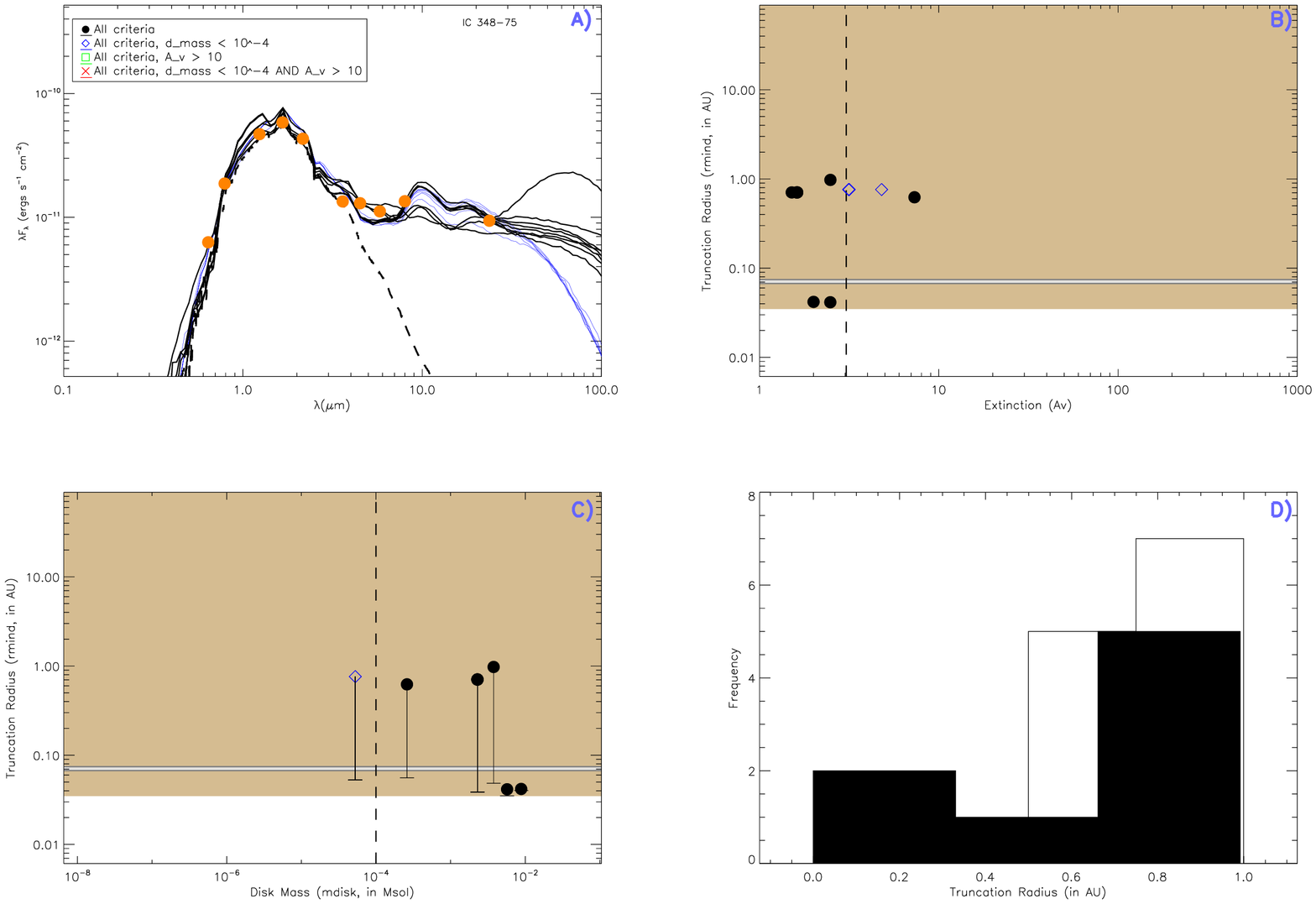}
\caption{IC 348 75. Refer to Figure \ref{ic348-36} for further explanation of each of the above panels. Figures 7 - 37 are available in the online version of the Journal.}
\label{ic348-75}
\end{figure}

\begin{figure}
\figurenum{17}
\centering
\includegraphics[angle=90, scale=0.75]{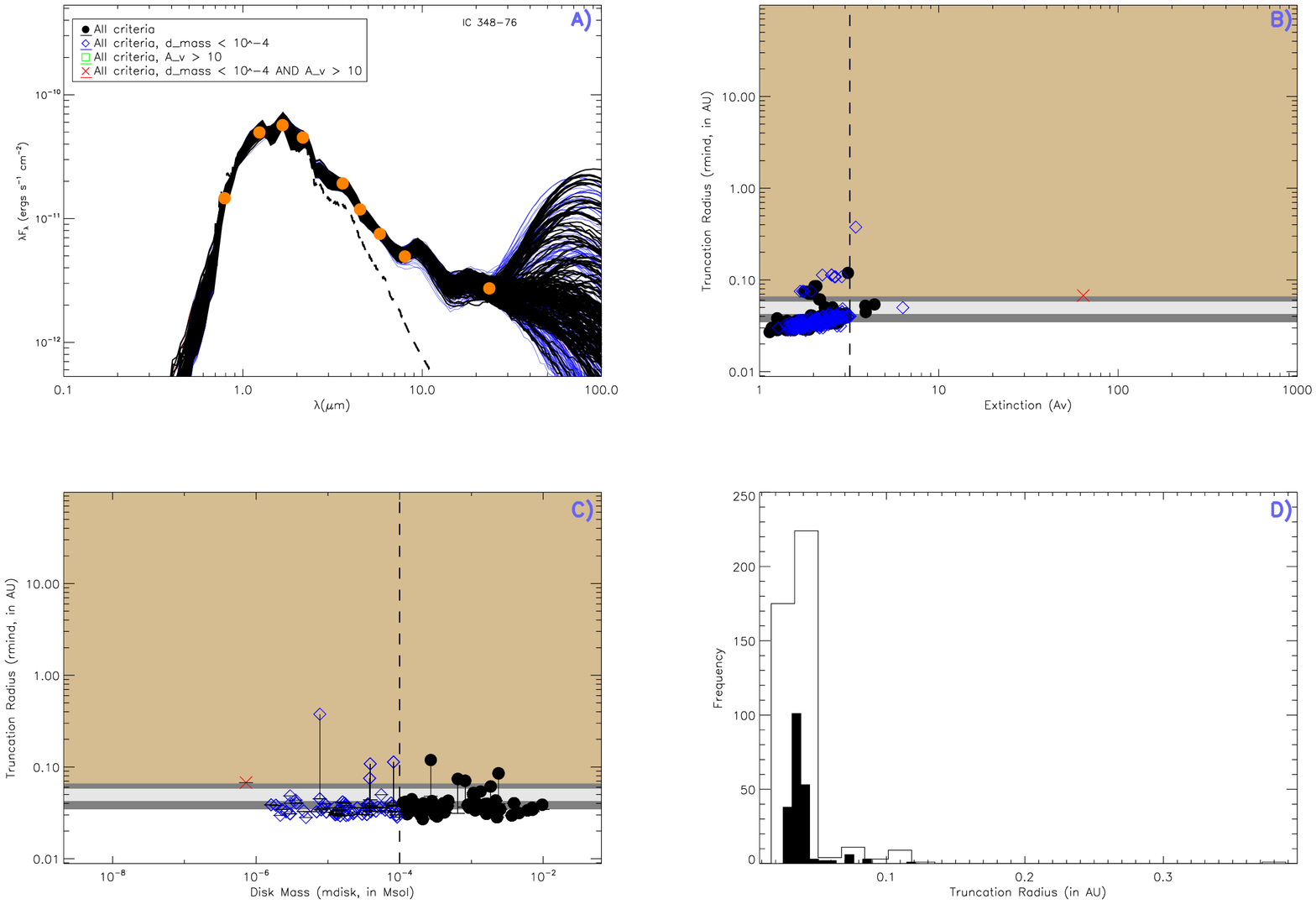}
\caption{IC 348 76. Refer to Figure \ref{ic348-36} for further explanation of each of the above panels. Figures 7 - 37 are available in the online version of the Journal.}
\label{ic348-76}
\end{figure}

\begin{figure}
\figurenum{18}
\centering
\includegraphics[angle=90, scale=0.75]{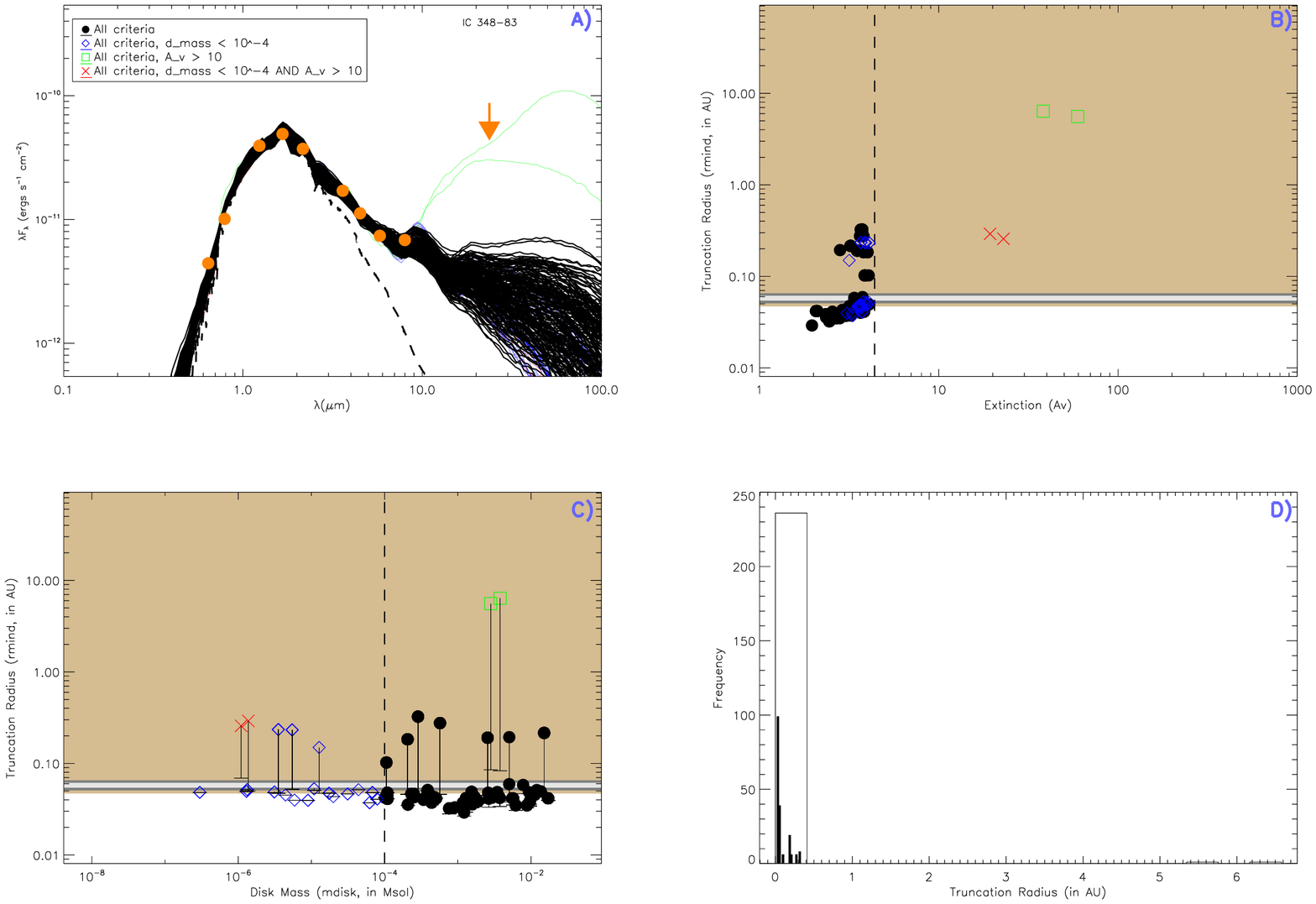}
\caption{IC 348 83. Refer to Figure \ref{ic348-36} for further explanation of each of the above panels. Figures 7 - 37 are available in the online version of the Journal.}
\label{ic348-83}
\end{figure}

\begin{figure}
\figurenum{19}
\centering
\includegraphics[angle=90, scale=0.75]{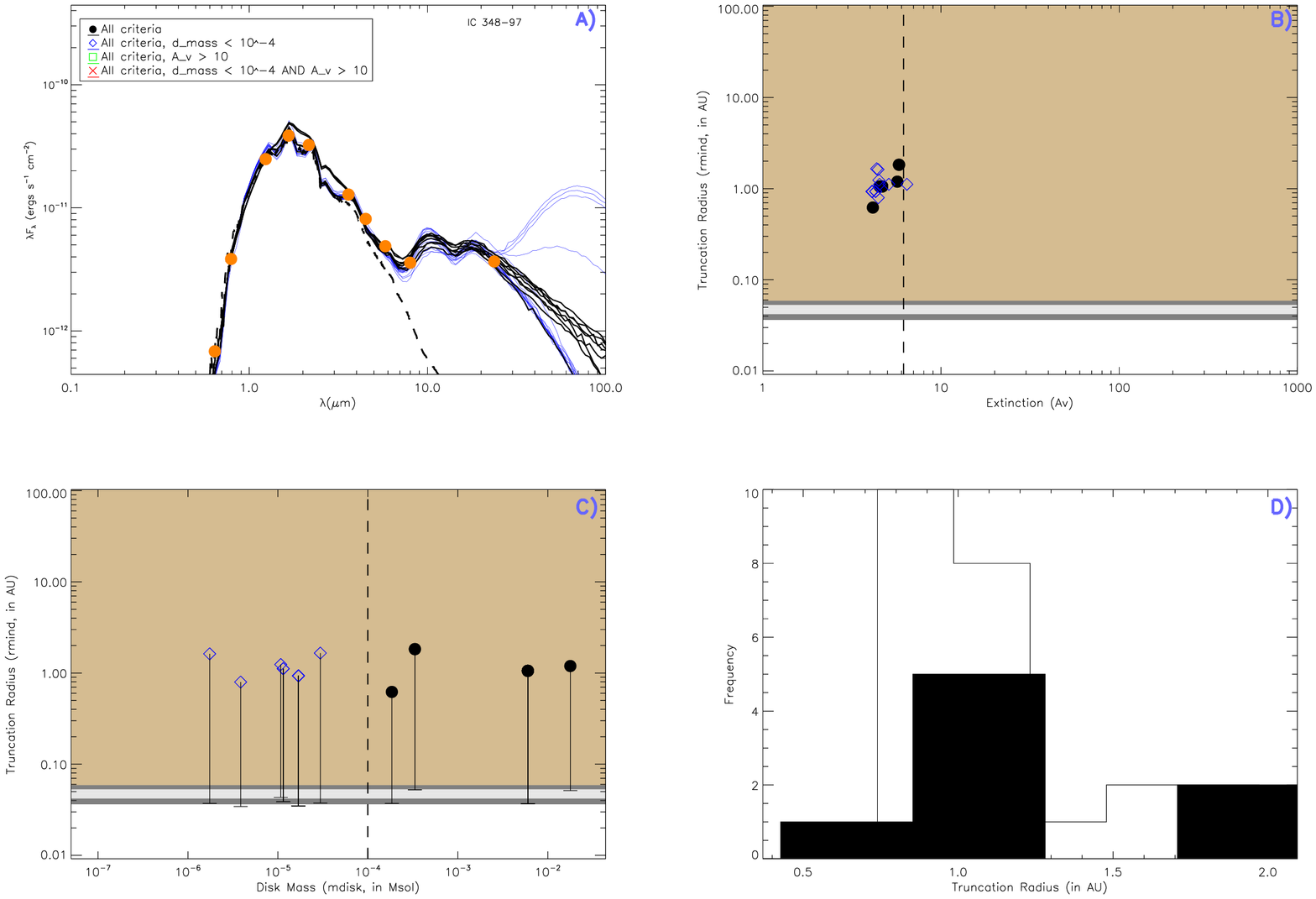}
\caption{IC 348 97. Refer to Figure \ref{ic348-36} for further explanation of each of the above panels. Figures 7 - 37 are available in the online version of the Journal.}
\label{ic348-97}
\end{figure}

\begin{figure}
\figurenum{20}
\centering
\includegraphics[angle=90, scale=0.75]{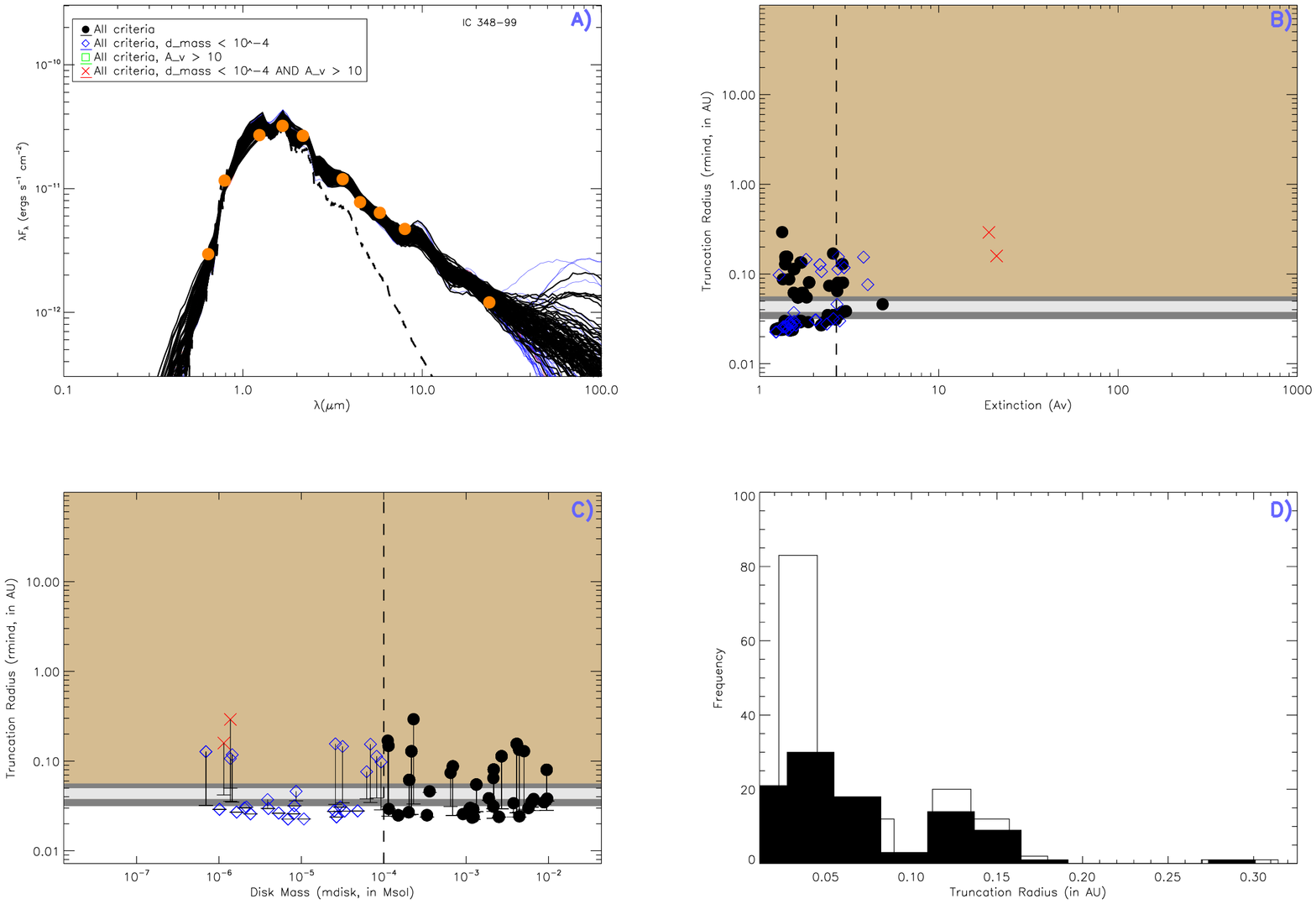}
\caption{IC 348 99. Refer to Figure \ref{ic348-36} for further explanation of each of the above panels. Figures 7 - 37 are available in the online version of the Journal.}
\label{ic348-99}
\end{figure}

\begin{figure}
\figurenum{21}
\centering
\includegraphics[angle=90, scale=0.75]{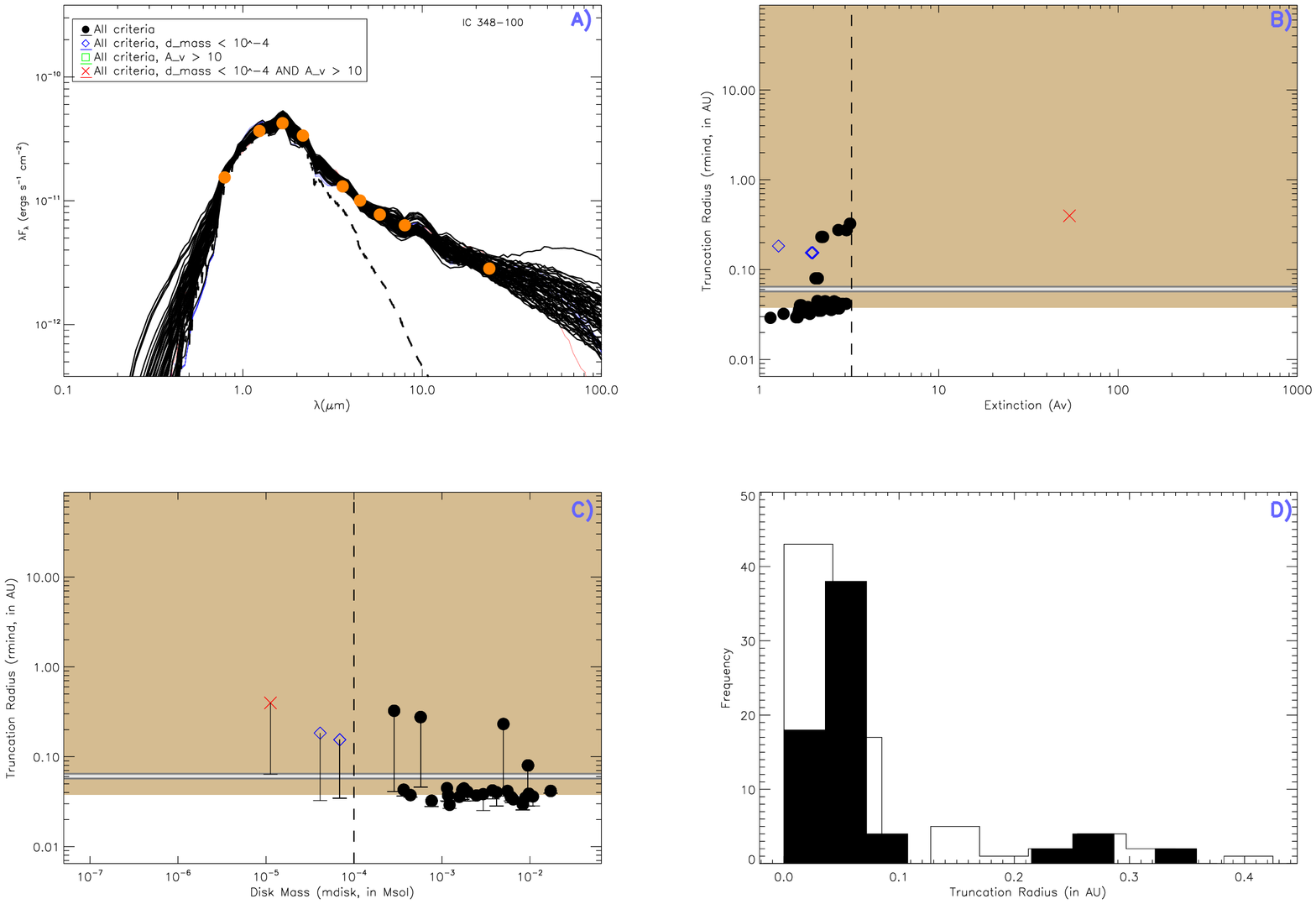}
\caption{IC 348 100. Refer to Figure \ref{ic348-36} for further explanation of each of the above panels. Figures 7 - 37 are available in the online version of the Journal.}
\label{ic348-100}
\end{figure}

\begin{figure}
\figurenum{22}
\centering
\includegraphics[angle=90, scale=0.75]{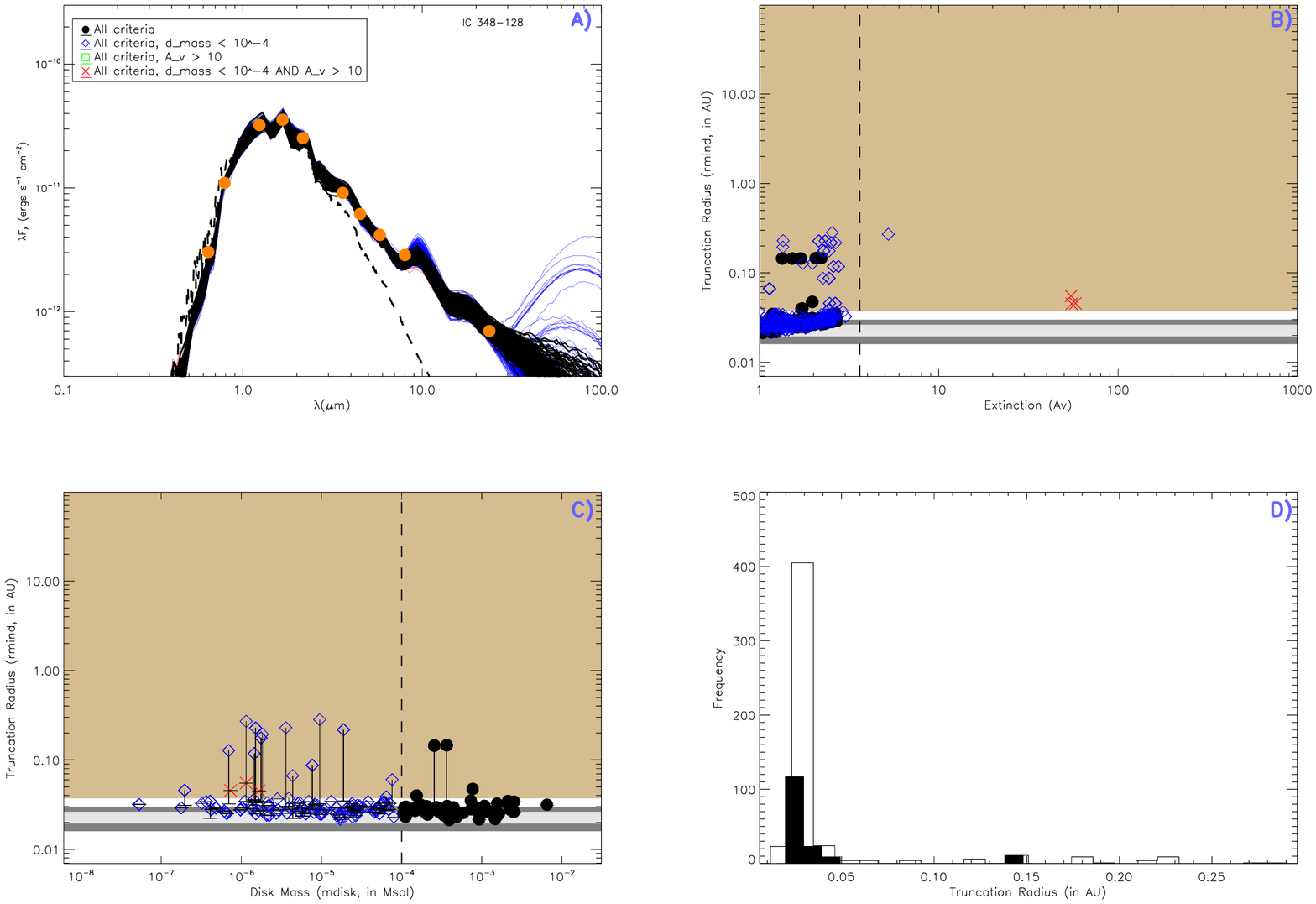}
\caption{IC 348 128. Refer to Figure \ref{ic348-36} for further explanation of each of the above panels. Figures 7 - 37 are available in the online version of the Journal.}
\label{ic348-6128}
\end{figure}

\begin{figure}
\figurenum{23}
\centering
\includegraphics[angle=90, scale=0.75]{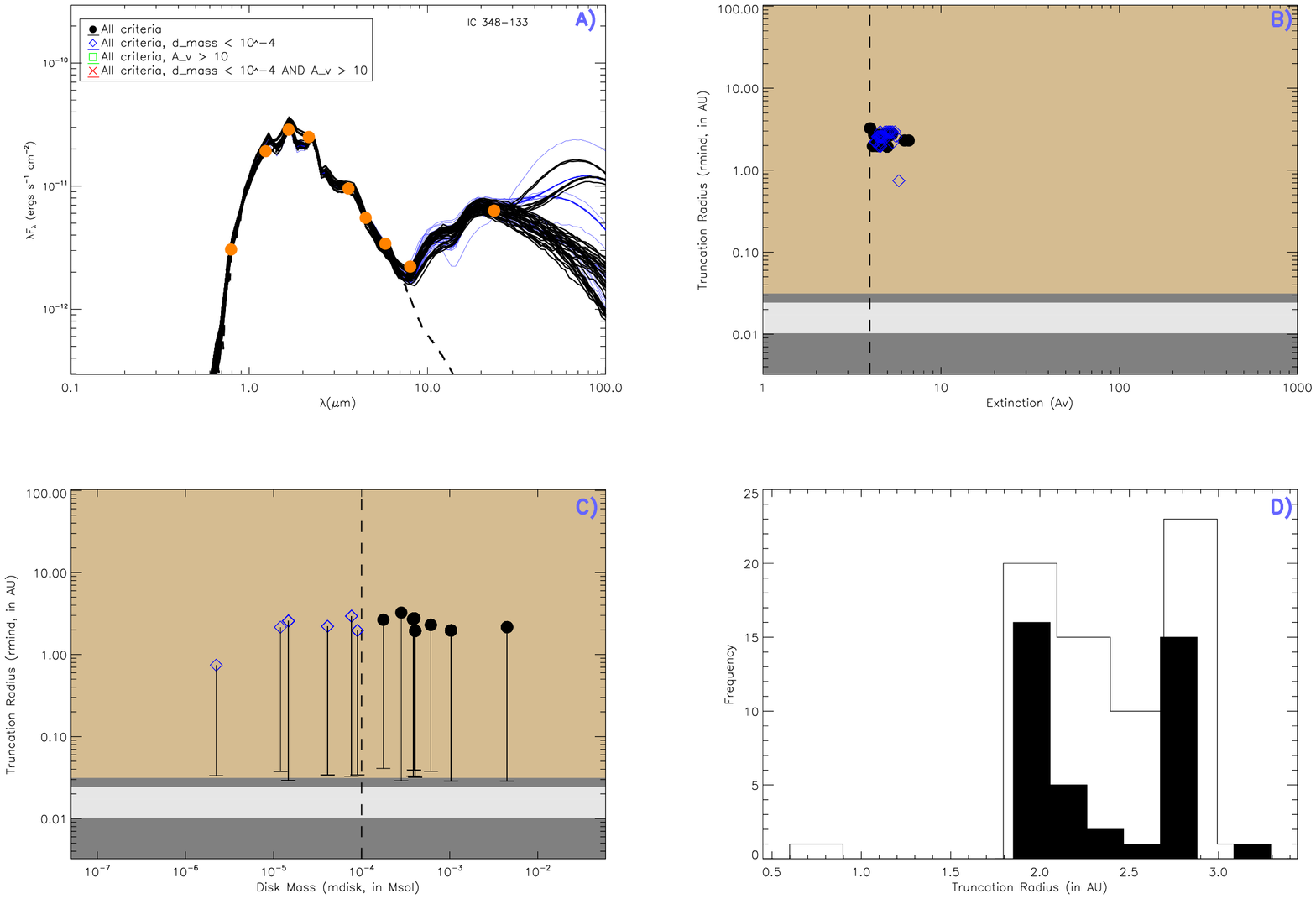}
\caption{IC 348 133. Refer to Figure \ref{ic348-36} for further explanation of each of the above panels. Figures 7 - 37 are available in the online version of the Journal.}
\label{ic348-133}
\end{figure}

\begin{figure}
\figurenum{24}
\centering
\includegraphics[angle=90, scale=0.75]{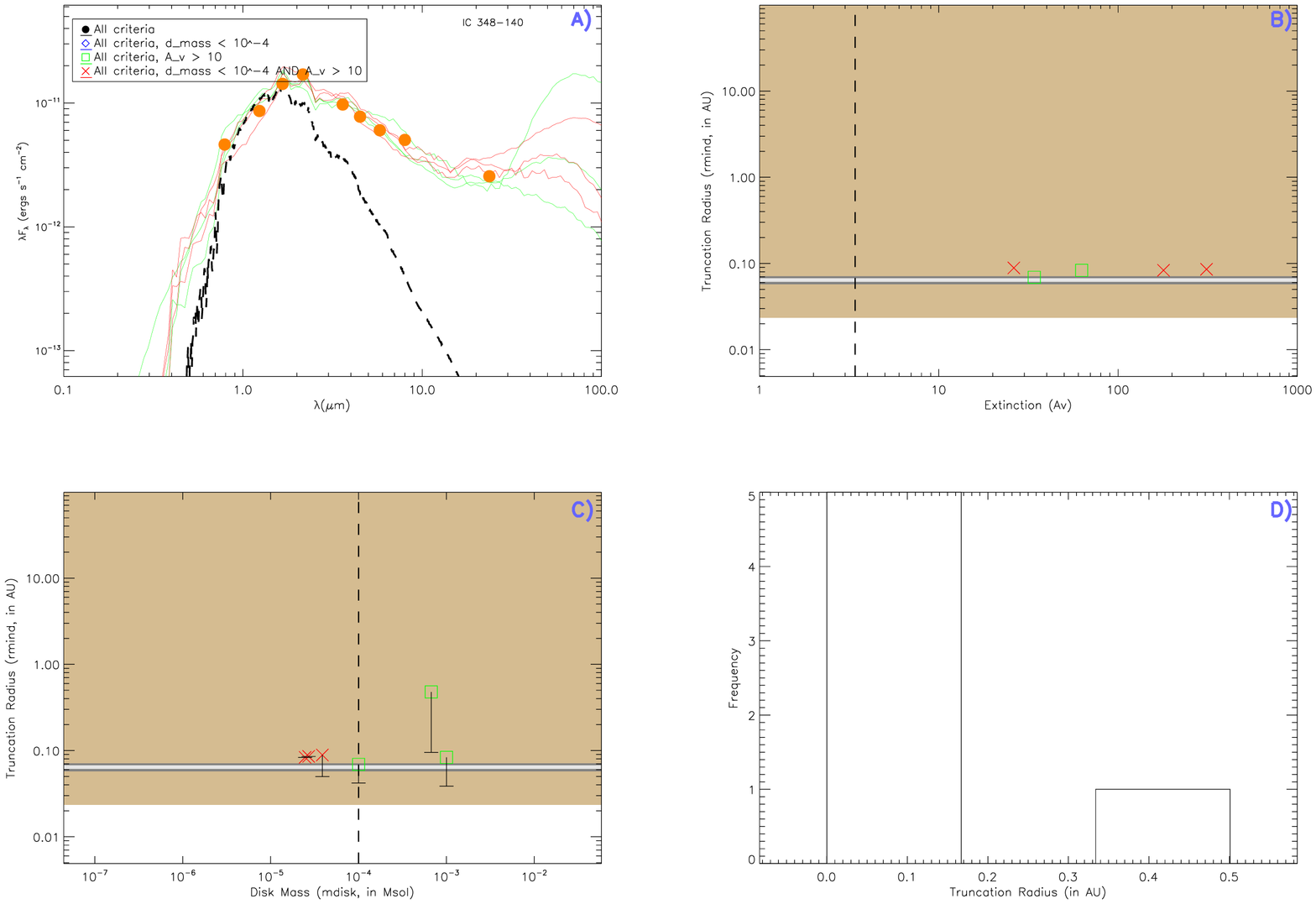}
\caption{IC 348 140. Refer to Figure \ref{ic348-36} for further explanation of each of the above panels. Figures 7 - 37 are available in the online version of the Journal.}
\label{ic348-140}
\end{figure}

\begin{figure}
\figurenum{25}
\centering
\includegraphics[angle=90, scale=0.75]{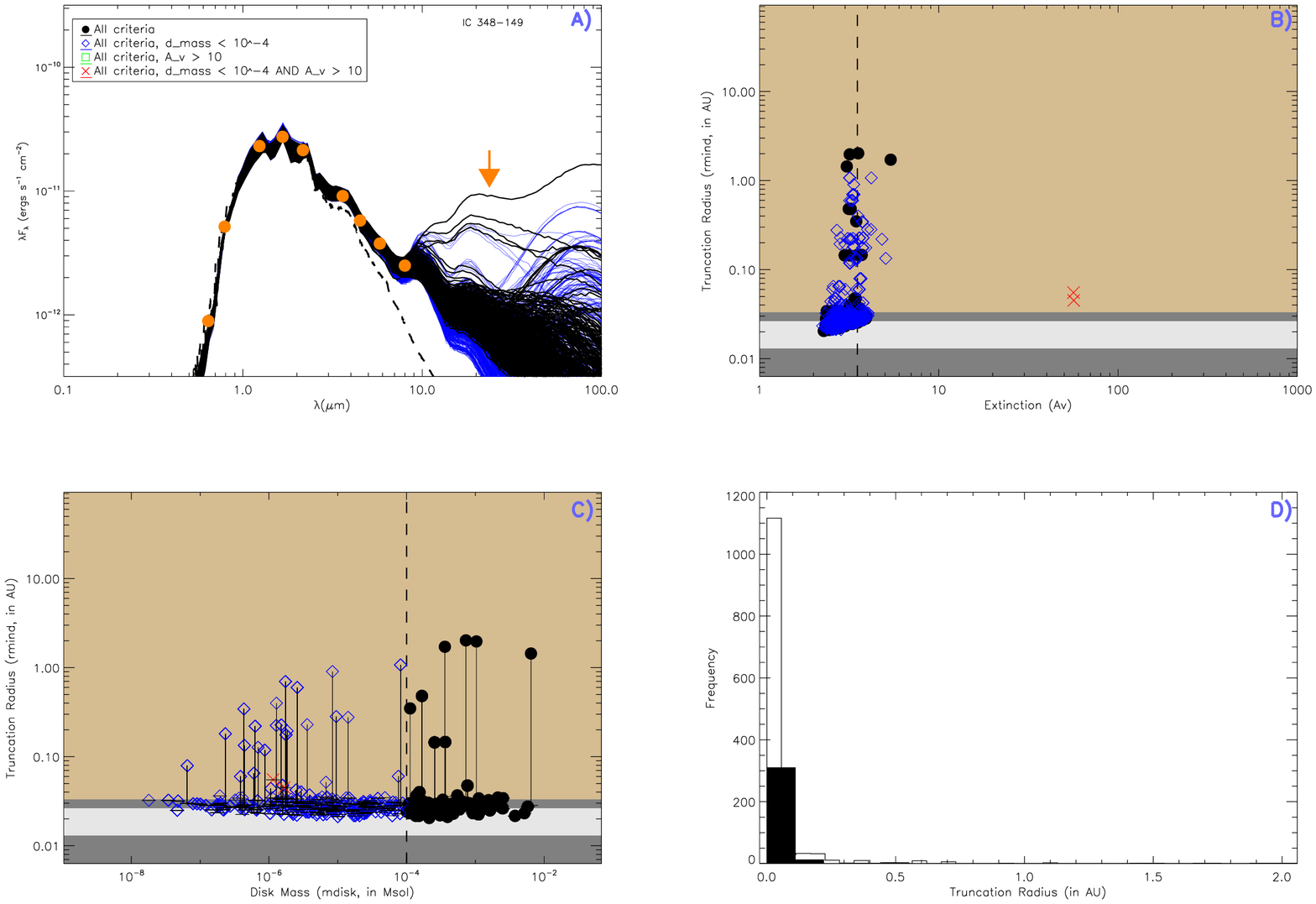}
\caption{IC 348 149. Refer to Figure \ref{ic348-36} for further explanation of each of the above panels. Figures 7 - 37 are available in the online version of the Journal.}
\label{ic348-149}
\end{figure}

\begin{figure}
\figurenum{26}
\centering
\includegraphics[angle=90, scale=0.75]{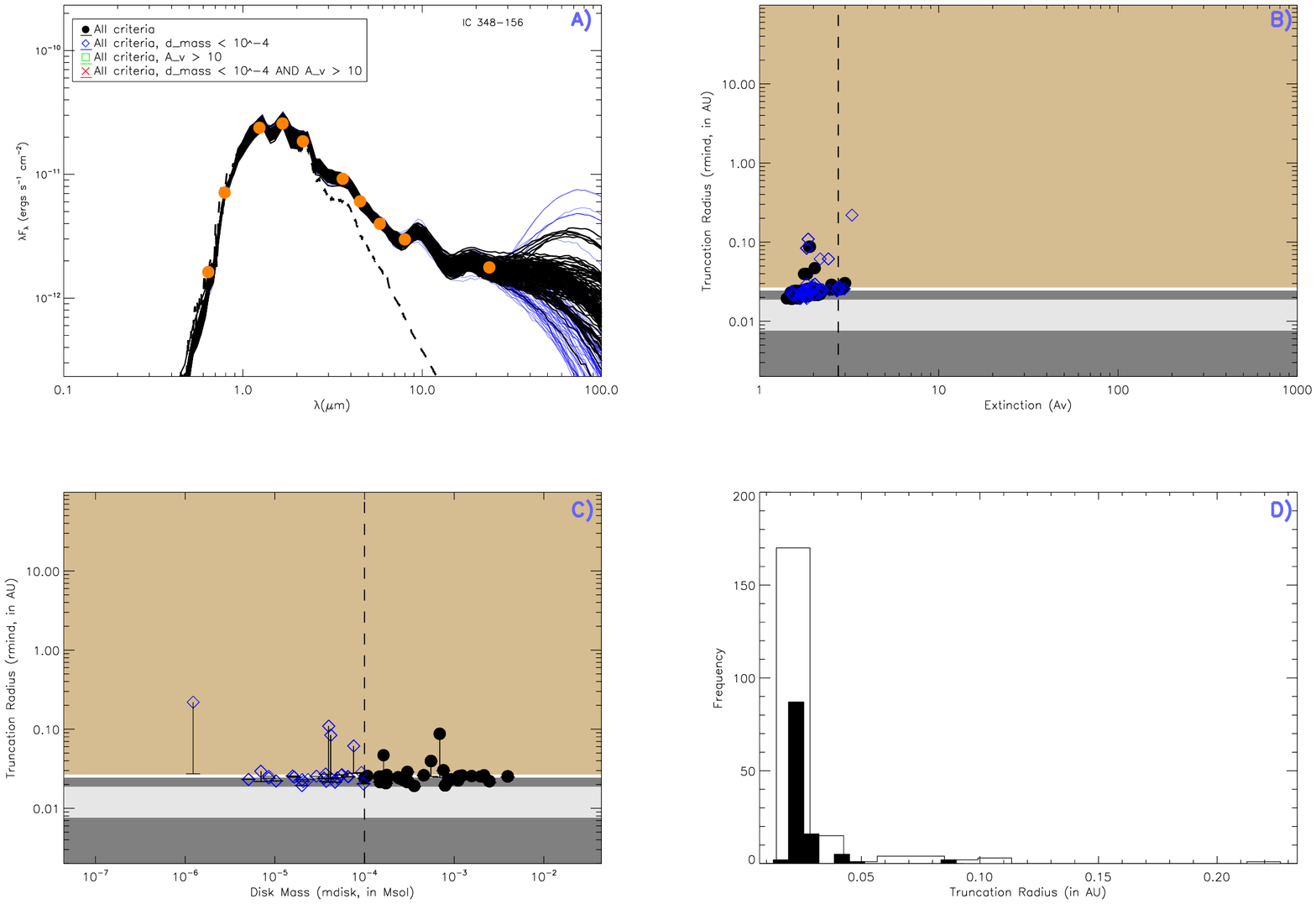}
\caption{IC 348 156. Refer to Figure \ref{ic348-36} for further explanation of each of the above panels. Figures 7 - 37 are available in the online version of the Journal.}
\label{ic348-156}
\end{figure}

\begin{figure}
\figurenum{27}
\centering
\includegraphics[angle=90, scale=0.75]{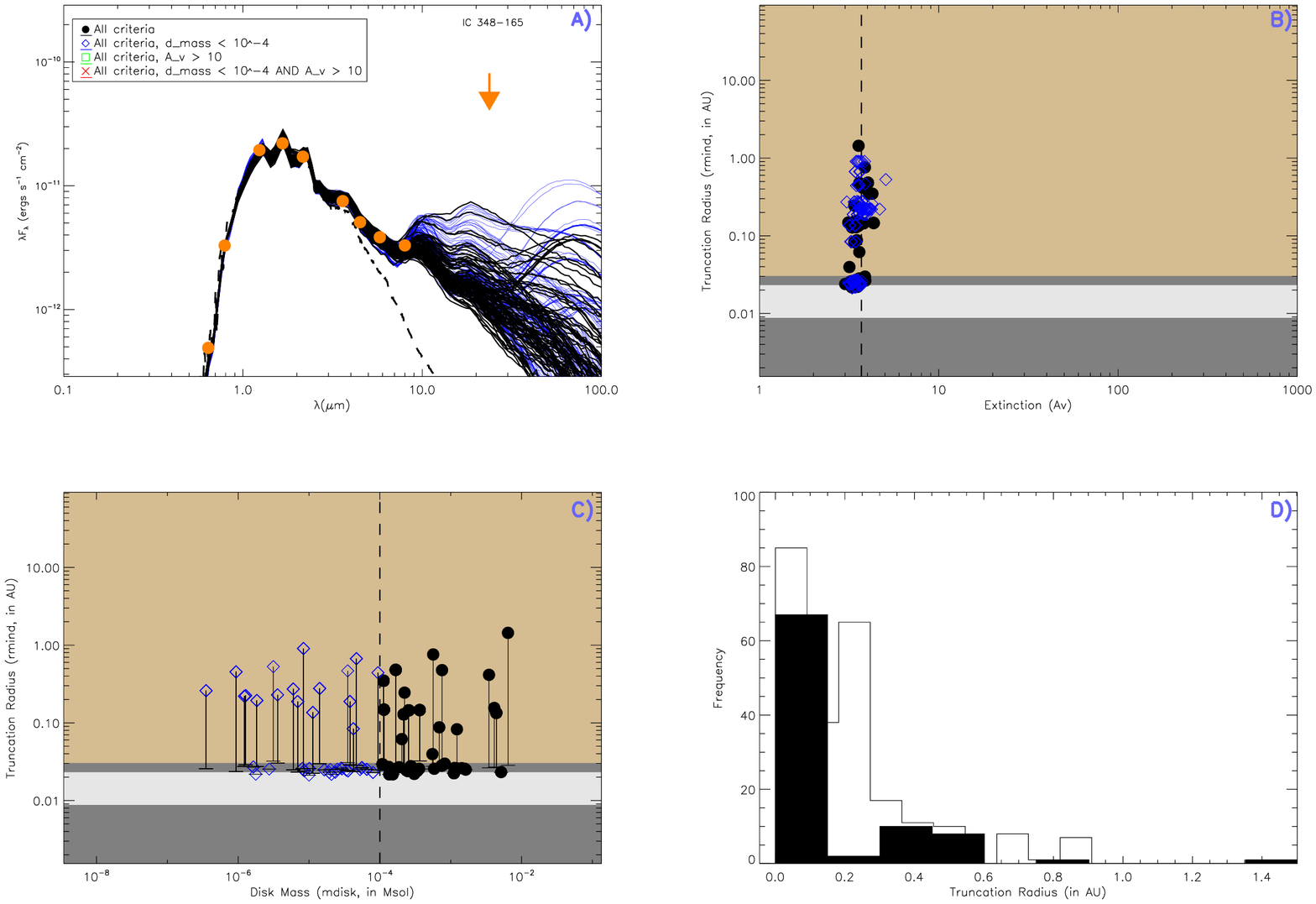}
\caption{IC 348 165. Refer to Figure \ref{ic348-36} for further explanation of each of the above panels. Figures 7 - 37 are available in the online version of the Journal.}
\label{ic348-165}
\end{figure}

\begin{figure}
\figurenum{28}
\centering
\includegraphics[angle=90, scale=0.75]{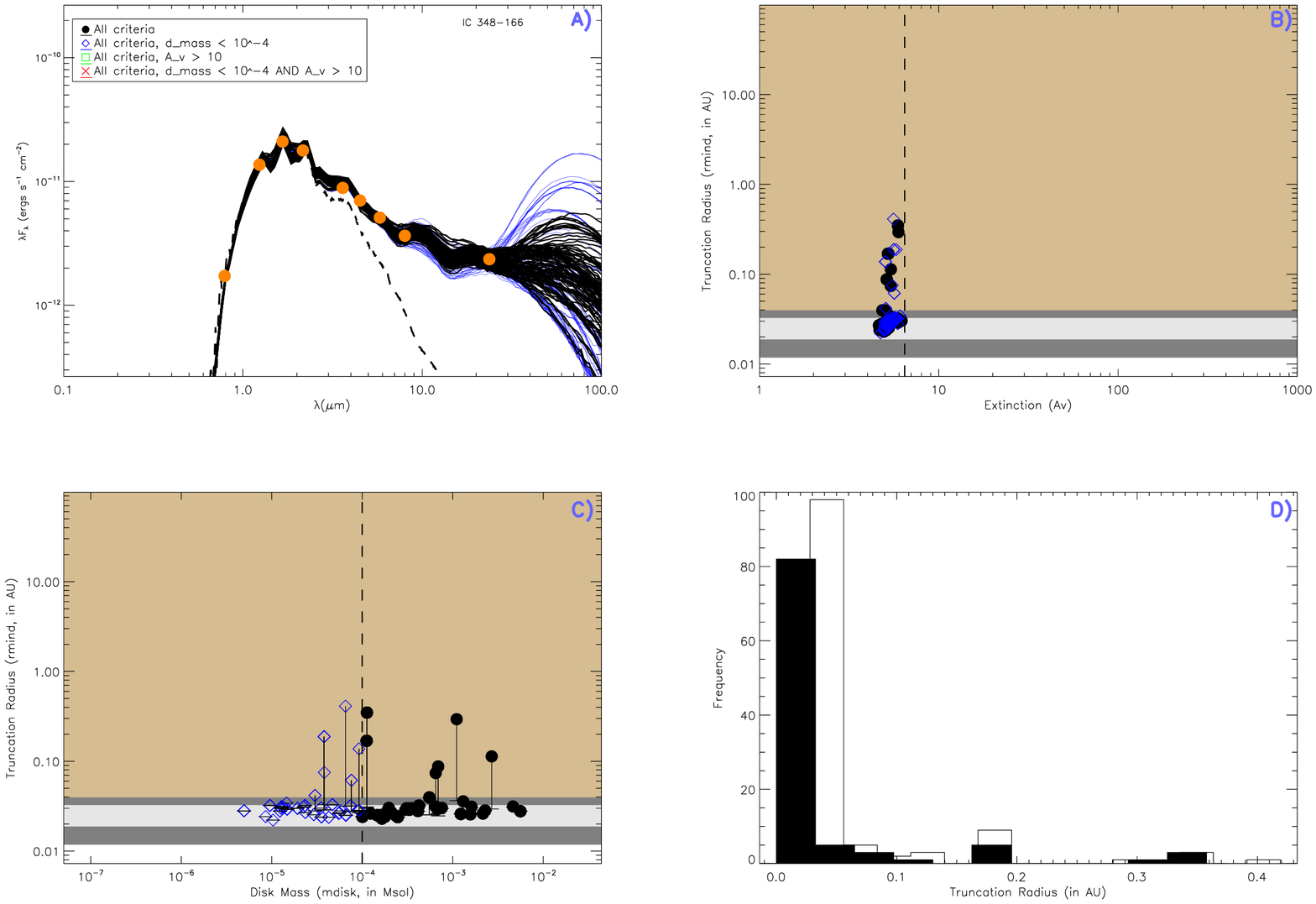}
\caption{IC 348 166. Refer to Figure \ref{ic348-36} for further explanation of each of the above panels. Figures 7 - 37 are available in the online version of the Journal.}
\label{ic348-166}
\end{figure}

\begin{figure}
\figurenum{29}
\centering
\includegraphics[angle=90, scale=0.75]{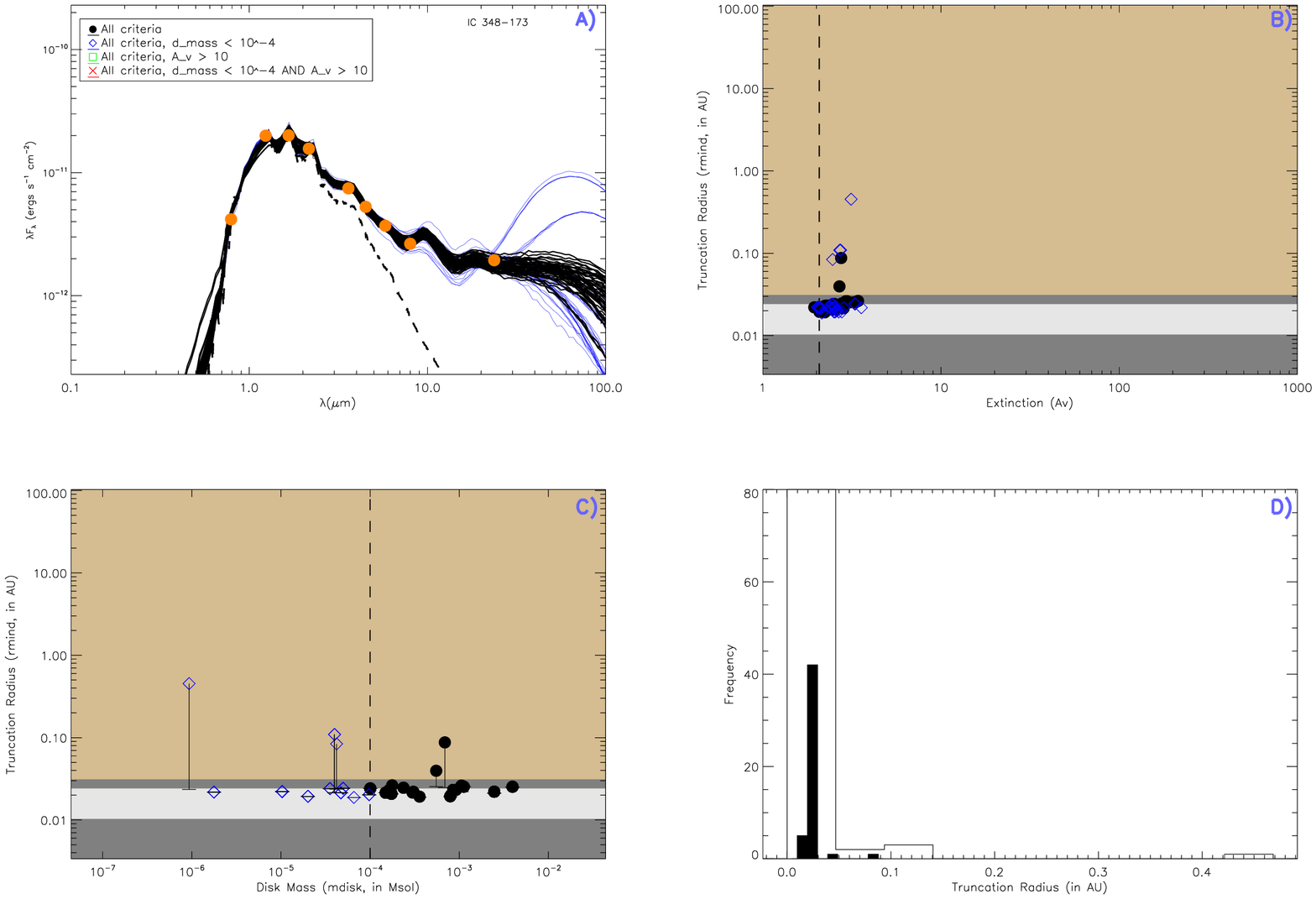}
\caption{IC 348 173. Refer to Figure \ref{ic348-36} for further explanation of each of the above panels. Figures 7 - 37 are available in the online version of the Journal.}
\label{ic348-173}
\end{figure}

\begin{figure}
\figurenum{30}
\centering
\includegraphics[angle=90, scale=0.75]{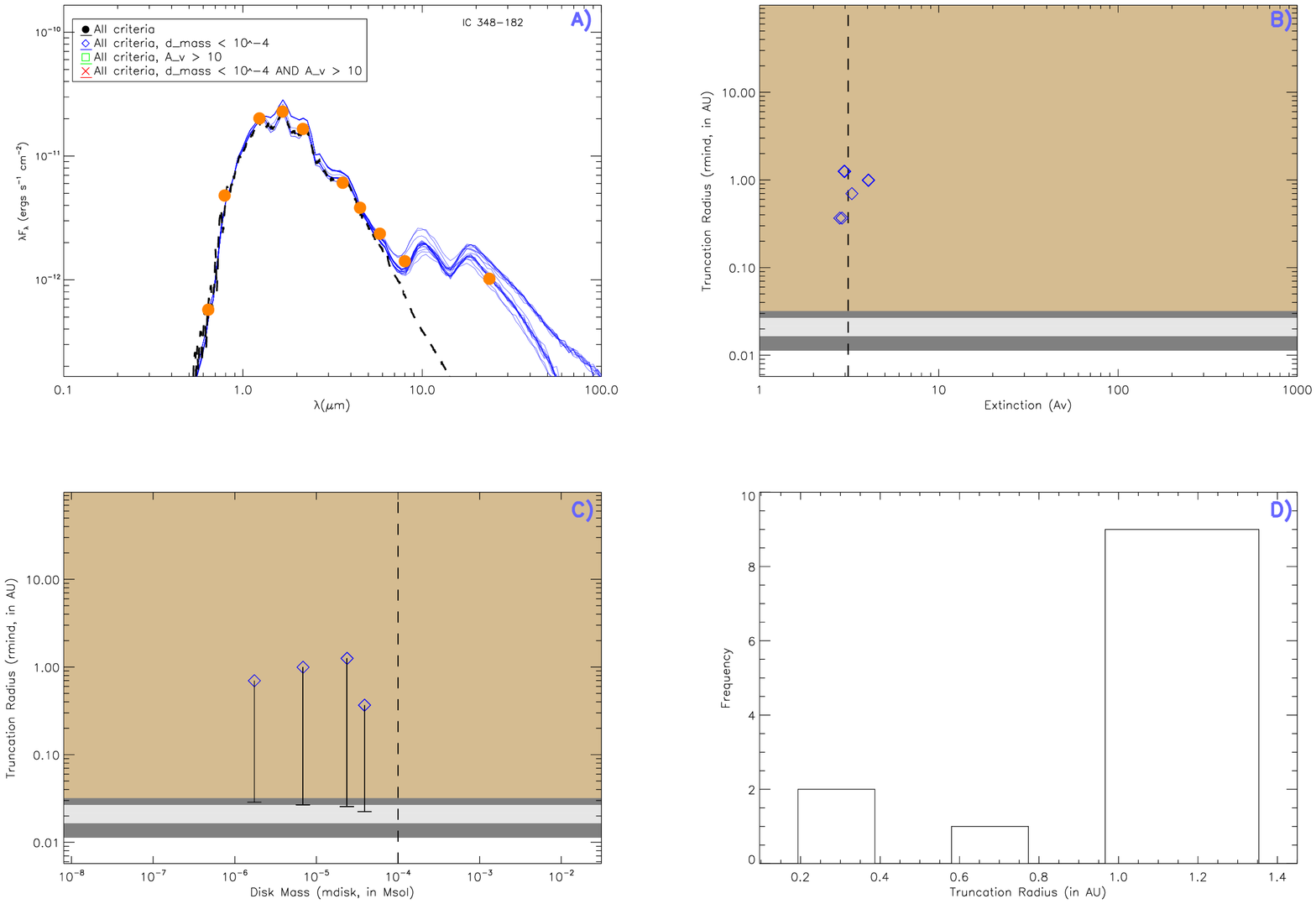}
\caption{IC 348 182. Refer to Figure \ref{ic348-36} for further explanation of each of the above panels. Figures 7 - 37 are available in the online version of the Journal.}
\label{ic348-182}
\end{figure}

\begin{figure}
\figurenum{31}
\centering
\includegraphics[angle=90, scale=0.75]{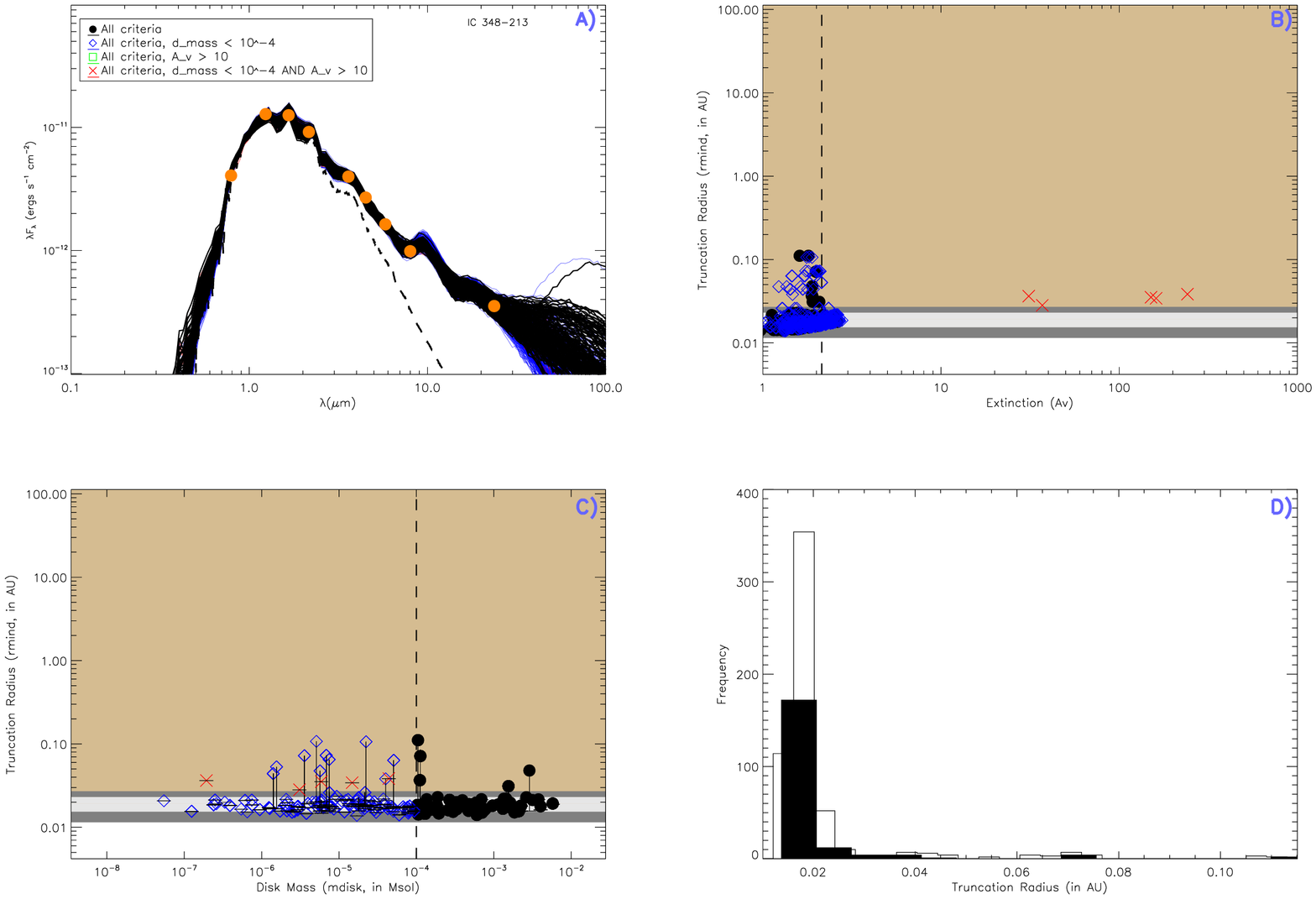}
\caption{IC 348 213. Refer to Figure \ref{ic348-36} for further explanation of each of the above panels. Figures 7 - 37 are available in the online version of the Journal.}
\label{ic348-213}
\end{figure}

\begin{figure}
\figurenum{32}
\centering
\includegraphics[angle=90, scale=0.75]{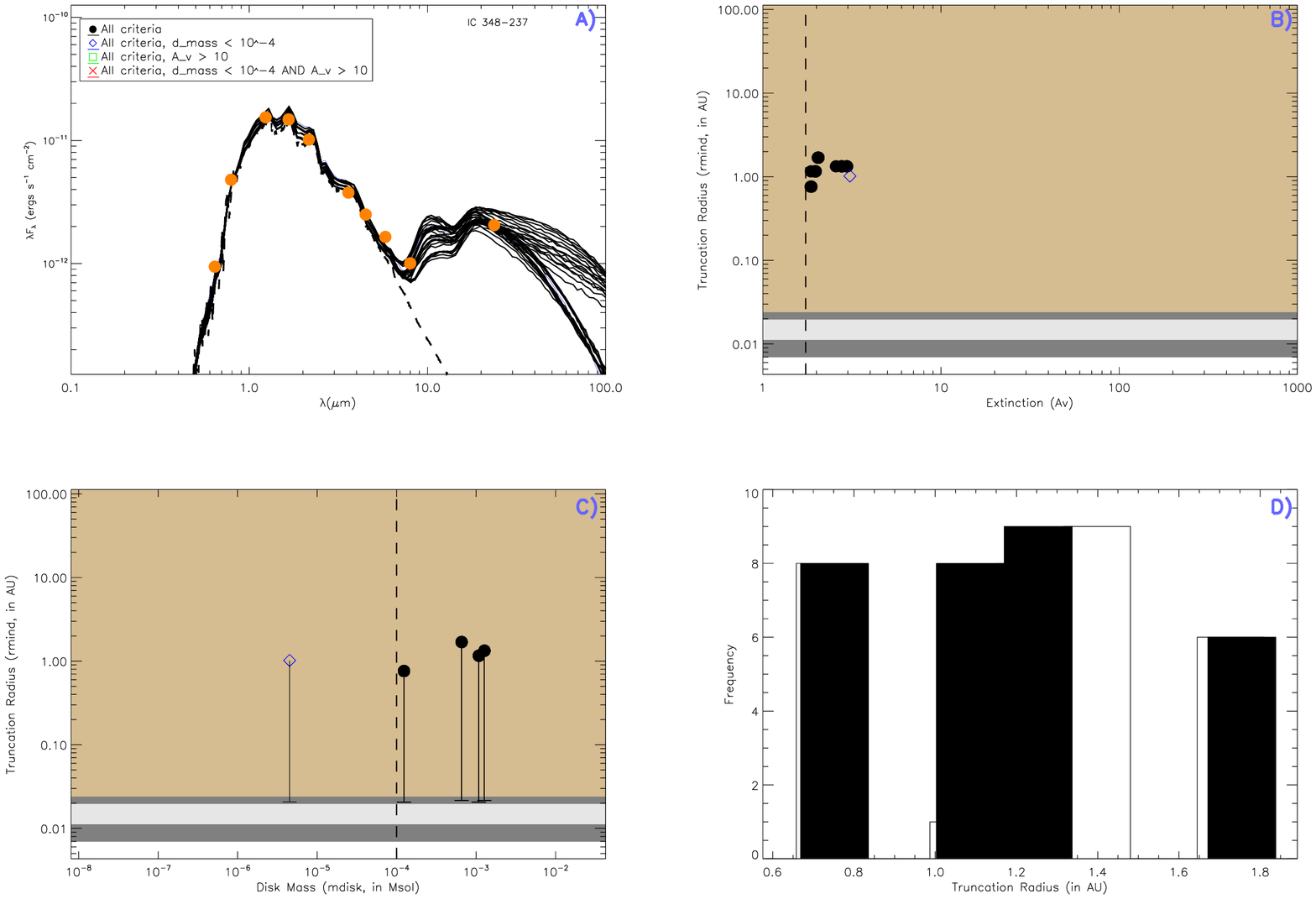}
\caption{IC 348 237. Refer to Figure \ref{ic348-36} for further explanation of each of the above panels. Figures 7 - 37 are available in the online version of the Journal.}
\label{ic348-237}
\end{figure}

\begin{figure}
\figurenum{33}
\centering
\includegraphics[angle=90, scale=0.75]{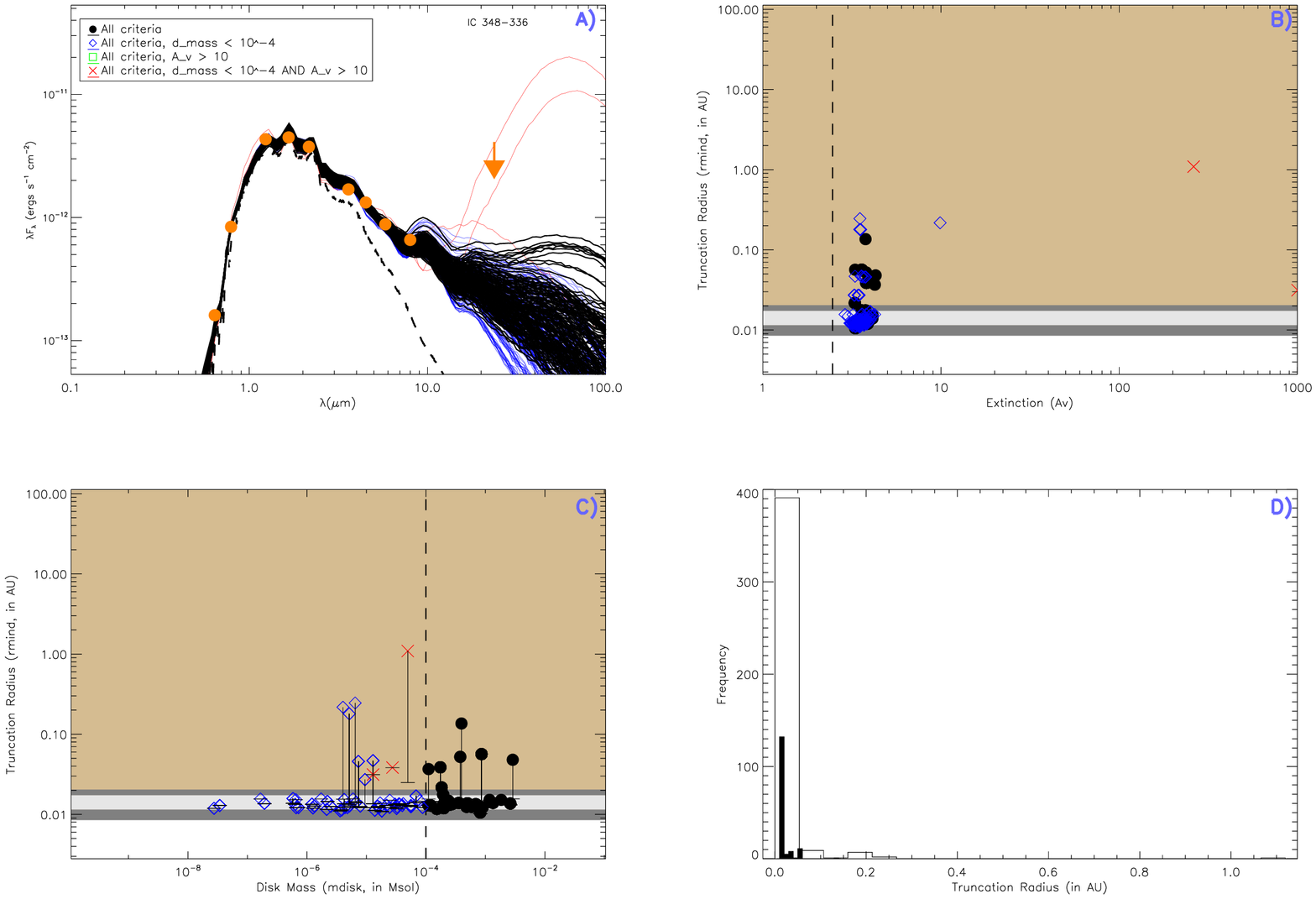}
\caption{IC 348 336. Refer to Figure \ref{ic348-36} for further explanation of each of the above panels. Figures 7 - 37 are available in the online version of the Journal.}
\label{ic348-336}
\end{figure}

\begin{figure}
\figurenum{34}
\centering
\includegraphics[angle=90, scale=0.75]{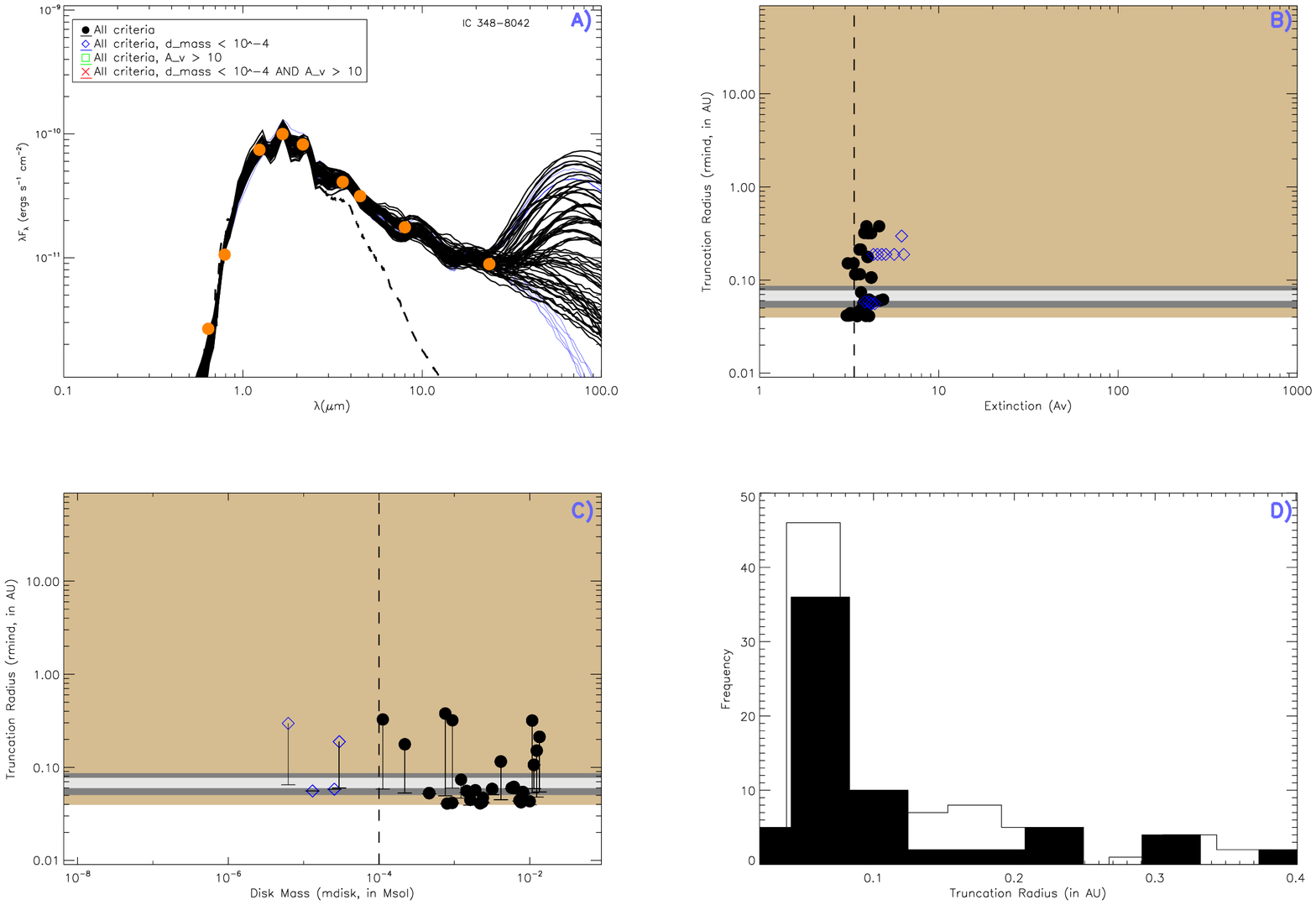}
\caption{IC 348 8042. Refer to Figure \ref{ic348-36} for further explanation of each of the above panels. Figures 7 - 37 are available in the online version of the Journal.}
\label{ic348-8042}
\end{figure}

\begin{figure}
\figurenum{35}
\centering
\includegraphics[angle=90, scale=0.75]{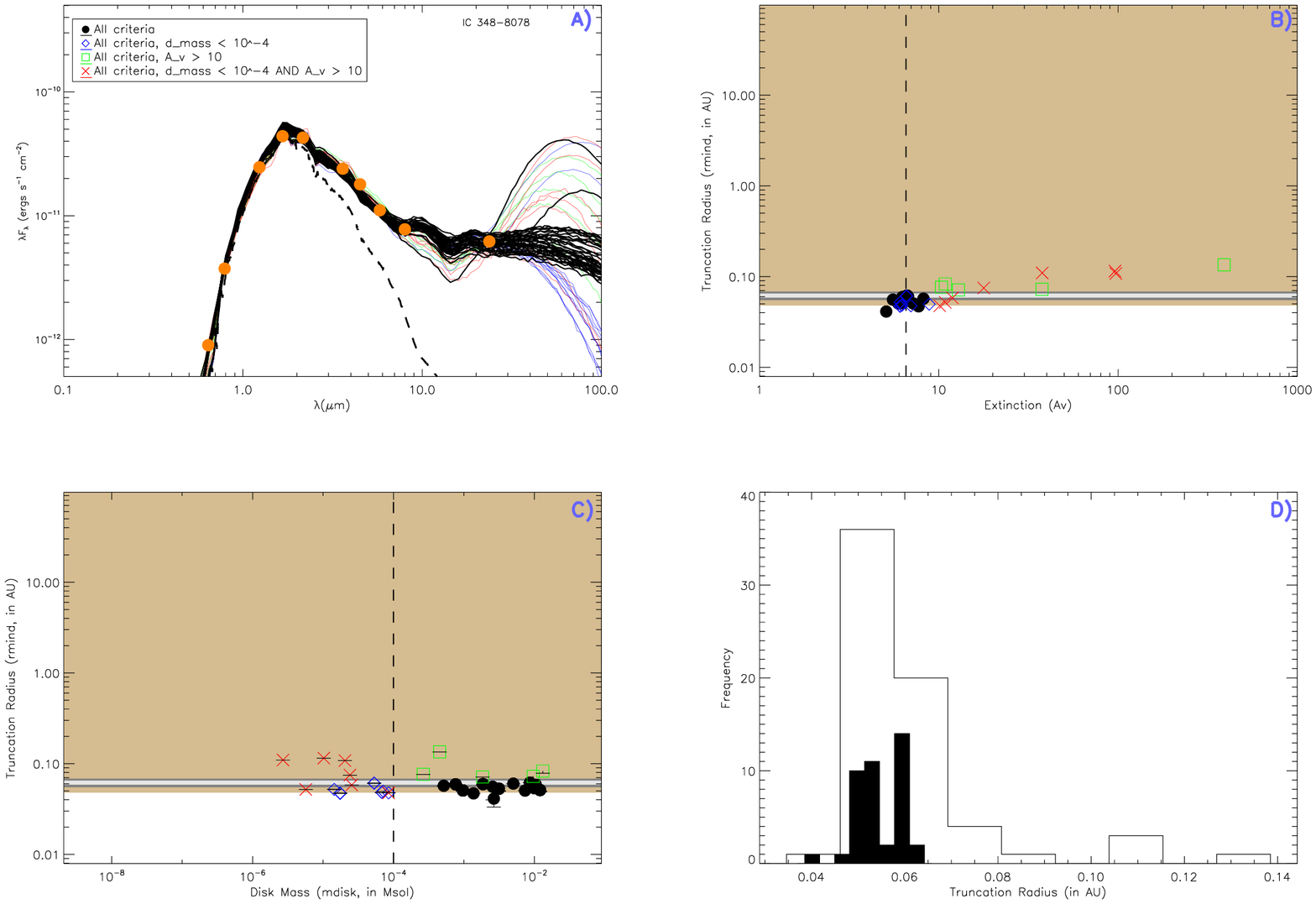}
\caption{IC 348 8078. Refer to Figure \ref{ic348-36} for further explanation of each of the above panels. Figures 7 - 37 are available in the online version of the Journal.}
\label{ic348-8078}
\end{figure}

\begin{figure}
\figurenum{36}
\centering
\includegraphics[angle=90, scale=0.75]{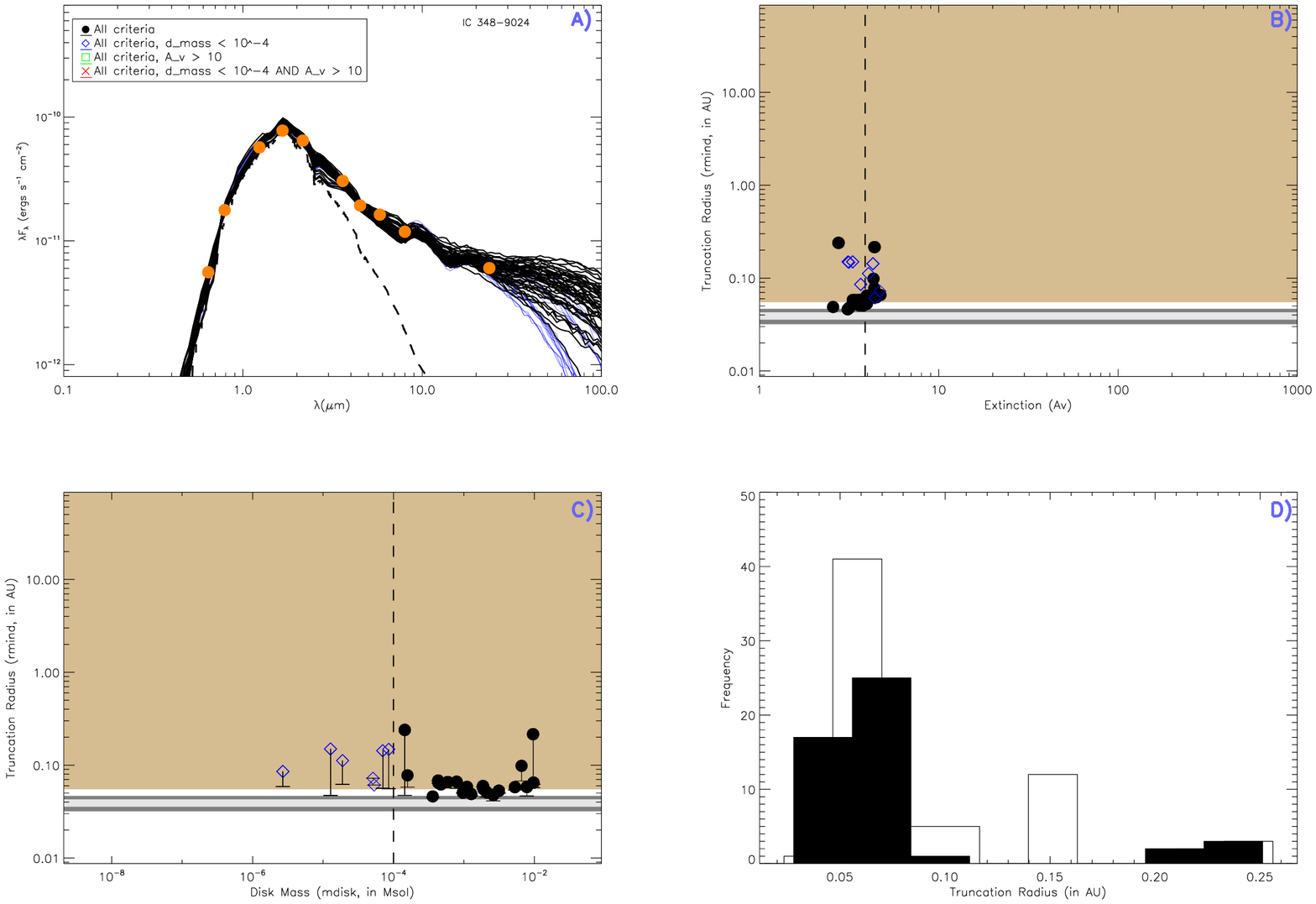}
\caption{IC 348 9024. Refer to Figure \ref{ic348-36} for further explanation of each of the above panels. Figures 7 - 37 are available in the online version of the Journal.}
\label{ic348-9024}
\end{figure}


\begin{figure}
\figurenum{37}
\centering
\includegraphics[angle=90, scale=0.75]{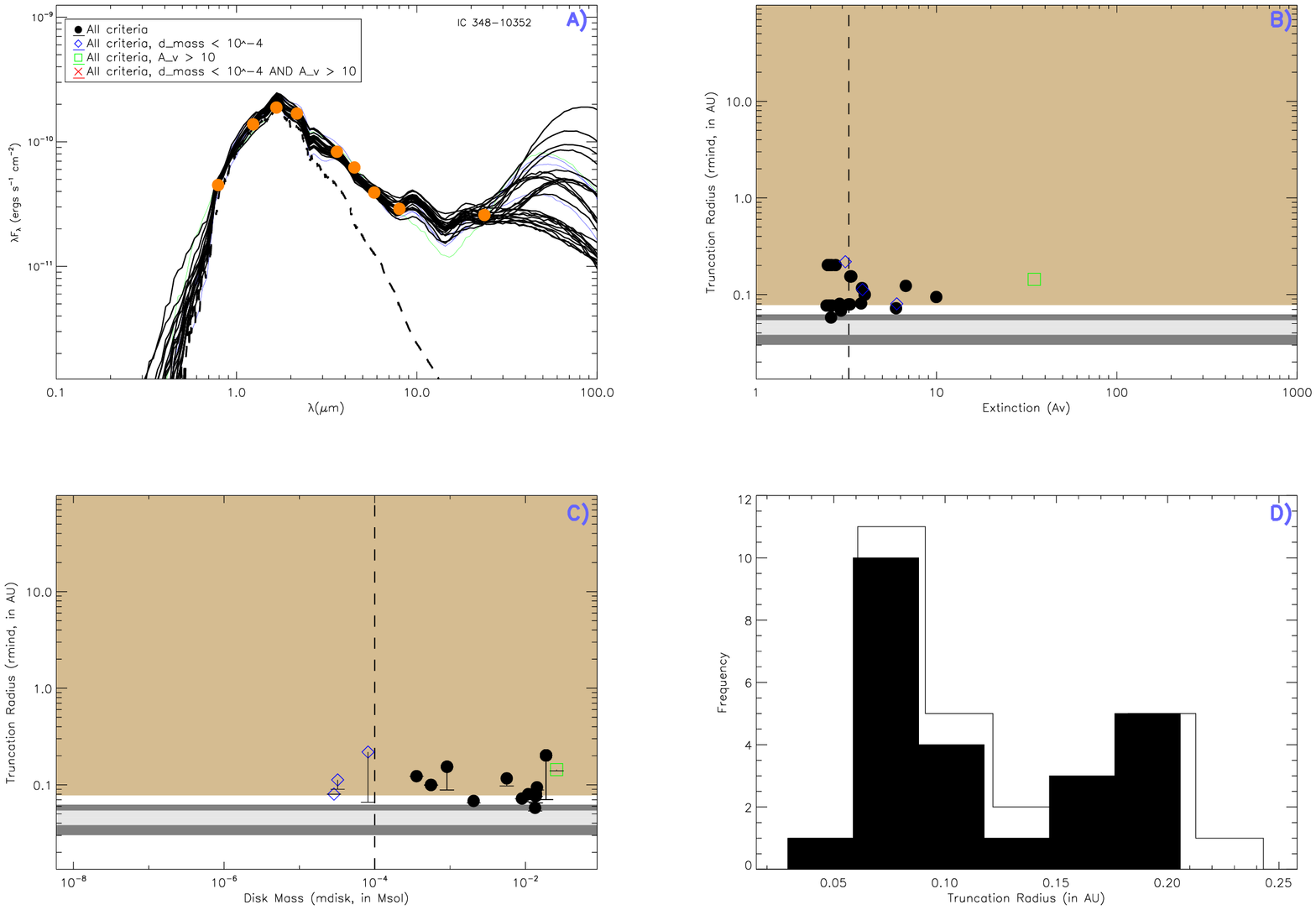}
\caption{IC 348 10352. Refer to Figure \ref{ic348-36} for further explanation of each of the above panels. Figures 7 - 37 are available in the online version of the Journal.}
\label{ic348-10352}
\end{figure}


\begin{deluxetable}{ccccc}
  \tablecaption{Photometric Calibration Data}
  \tablewidth{0pt}
  \tabletypesize{\footnotesize}
  \tablehead{
  \colhead{Filter(Type)} & \colhead{$\lambda_{eff}(\mu m)$} & \colhead{Zeropoint (Jy)} & \colhead{Refs.}
  }
  \startdata
Johnson[R]	&	0.64	&	3072.0 	&	\tablenotemark{1}\\
Johnson[I]		&	0.79	&	2496.4 	&	\tablenotemark{1}\\
2MASS[J]		&	1.24	&	1594.0 	&	\tablenotemark{2}\\
2MASS[H]		&	1.65	&	1024.0 	&	\tablenotemark{2}\\
2MASS[K]		&	2.17	&	666.7 	&	\tablenotemark{2}\\
IRAC[3.6]		&	3.6	&	277.3	&	\tablenotemark{3}\\
IRAC[4.5]		&	4.5	&	179.6	&	\tablenotemark{3}\\
IRAC[5.8]		&	5.8	&	116.6	&	\tablenotemark{3}\\
IRAC[8.0]		&	8.0	&	63.1		&	\tablenotemark{3}\\
MIPS[24]		&	24.0	&	7.14		&	\tablenotemark{3}
  \enddata
  \tablecomments{Calibration data used in converting photometric measurements into flux.}
  \tablenotetext{1}{\citet[]{Cousins:1976aa}}
  \tablenotetext{2}{\citet[]{Cohen:2003aa}}
  \tablenotetext{3}{\citet[]{Fazio:2004aa}}
  \label{calibrationtable}
\end{deluxetable}

\begin{deluxetable}{ccccccc|ccc}
  \tablecaption{Derived Stellar and Disk Parameters of IC 348 Sample}
  \tablewidth{0pt}
  \tabletypesize{\scriptsize}
  \tablehead{
  \colhead{} & \multicolumn{6}{c}{Stellar Parameters {from the literature}} & \multicolumn{3}{c}{Disk Radii {from SED fitting}} \\
  \colhead{Star ID} & \colhead{$P_{rot}$ (d)} & \colhead{$T_{eff}$ (K)} & \colhead{L$_{bol}$ ($L_{\sun}$)} & \colhead{M ($M_{\sun}$) \tablenotemark{\dag}} & \colhead{R ($R_{\sun}$) \tablenotemark{\dag}} & \colhead{$A_V$ (mag)} & \colhead{R$_{sub}$ (R$_\star$) \tablenotemark{\ddag}} & \colhead{R$_{co}$ (R$_\star$)} & \colhead{R$_{trunc}$ (R$_\star$)}
  }
  \startdata
32					&	8.2	&	4060	&	1.4	&	0.48	&	2.39	&	5.79	&	0.08	&	0.06	&	0.11	\\
36					&	5.1	&	4205	&	1.5	&	0.56	&	2.31	&	2.91	&	0.08	&	0.05	&	0.51	\\
37					&	8.6	&	4205	&	0.99	&	0.64	&	1.87	&	2.88	&	0.07	&	0.07	&	0.27	\\
58					&	7.3	&	3669	&	0.72	&	0.31	&	2.1	&	2.59	&	0.06	&	0.05	&	0.04	\\
71					&	6.7	&	3415	&	0.47	&	0.24	&	1.96	&	2.14	&	0.04	&	0.04	&	2.66	\\
75					&	10.6	&	3669	&	0.28	&	0.44	&	1.31	&	2.94	&	0.03	&	0.07	&	0.71	\\
76					&	9.5	&	3306	&	0.39	&	0.19	&	1.9	&	2.13	&	0.04	&	0.05	&	0.04	\\
83					&	8.4	&	3705	&	0.51	&	0.37	&	1.73	&	3.43	&	0.05	&	0.06	&	0.04	\\
91					&	3.9	&	3560	&	0.39	&	0.32	&	1.64	&	2.01 &	0.04	&	0.03	&	0.03 \\
97					&	7.3	&	3524	&	0.54	&	0.28	&	1.97	&	4.63	&	0.05	&	0.05	&	1.06	\\
99					&	7.6	&	3306	&	0.26	&	0.21	&	1.55	&	1.91	&	0.03	&	0.04	&	0.04	\\
100					&	8.4	&	3705	&	0.33	&	0.44	&	1.39	&	2.23	&	0.04	&	0.06	&	0.04	\\
110					&	19.8	&	3560	&	0.34	&	0.34	&	1.53	&	4.66	&	0.04	&	0.10	&	2.18	\\
128					&	2.2	&	3560	&	0.32	&	0.34	&	1.49	&	1.73	&	0.04	&	0.02	&	0.03	\\
133					&	2.1	&	3125	&	0.17	&	0.16	&	1.41	&	4.79	&	0.03	&	0.02	&	2.16	\\
149					&	2.5	&	3161	&	0.18	&	0.17	&	1.41	&	3.04	&	0.03	&	0.02	&	0.03	\\
156					&	1.3	&	3234	&	0.17	&	0.21	&	1.31	&	2.04	&	0.03	&	0.01	&	0.02	\\
165					&	1.9	&	3091	&	0.16	&	0.16	&	1.39	&	3.58	&	0.03	&	0.02	&	0.13	\\
166					&	3.6	&	3234	&	0.24	&	0.18	&	1.56	&	5.24	&	0.03	&	0.03	&	0.03	\\
173					&	2.2	&	3024	&	0.12	&	0.15	&	1.26	&	2.57	&	0.02	&	0.02	&	0.02	\\
213					&	2.3	&	3161	&	0.07	&	0.19	&	0.9	&	1.87	&	0.02	&	0.02	&	0.02	\\
237					&	1.7	&	3125	&	0.07	&	0.18	&	0.93	&	2.23	&	0.02	&	0.02	&	1.16	\\
336					&	1.6	&	3058	&	0.03	&	0.16	&	0.6	&	3.6	&	0.01	&	0.01	&	0.01	\\
8042					&	16	&	3234	&	0.37	&	0.17	&	1.94	&	3.97	&	0.04	&	0.07	&	0.06	\\
8078					&	8.9	&	3778	&	0.53	&	0.42	&	1.7	&	6.37	&	0.05	&	0.06	&	0.05	\\
9024					&	4.5	&	3850	&	0.69	&	0.41	&	1.87	&	3.82	&	0.05	&	0.04	&	0.06	\\
10352				&	6.9	&	3705	&	1.4	&	0.28	&	2.87	&	3.7	&	0.08	&	0.05	&	0.10	\\
LB06-100				&	19.8	&	3560	&	0.34	&	0.34	&	1.53	&	4.86 &	0.04	&	0.10	&	0.04 \\
\enddata
  \tablecomments{Derived stellar properties. The radius and mass measurements were derived as described in \S~\ref{stellarproperties}, while temperature, extinction, and luminosity values are from \citep{Luhman:2003aa}, and rotation periods from \citep{Cieza:2006aa}. Typical fractional errors for the following parameters are; $T_{eff} \sim$ 3\%, $L_{bol} \sim$ 81\%, $M_{\sun} \sim$ 61\%, and $R_{\sun} \sim$ 41\%.}
\tablenotetext{*}{These extinction values are the original estimates from L03. Our matches in this study for these stars did not yield enough information to make a judgement on the best fit extinction, and are excluded from the final analysis.}
\tablenotetext{\dag}{Masses and radii were estimated using DM97\citep{[]Dantona:1997aa} pre-main sequence evolutionary models (see \S\ref{stellarproperties}), and the Stefan-Boltzmann law, respectively.}
\label{derivedtable}
\tablenotetext{\ddag}{Sublimation radii were estimated using the stellar effective temperature and the sublimation temperature at which the dust in a protoplanetary disk is expected to be destroyed (see \S\ref{radii}).}
\end{deluxetable}

\clearpage

\begin{deluxetable}{ccccccc|ccc}
  \tablecaption{Discarded IC 348 Sample}
  \tablewidth{0pt}
  \tabletypesize{\scriptsize}
  \tablehead{
  \colhead{} & \multicolumn{6}{c}{Stellar Parameters {from the literature}} & \multicolumn{3}{c}{Disk Radii {from SED fitting}} \\
  \colhead{Star ID} & \colhead{$P_{rot}$ (d)} & \colhead{$T_{eff}$ (K)} & \colhead{L$_{bol}$ ($L_{\sun}$)} & \colhead{M ($M_{\sun}$) \tablenotemark{\dag}} & \colhead{R ($R_{\sun}$) \tablenotemark{\dag}} & \colhead{$A_V$ (mag)} & \colhead{R$_{sub}$ (R$_\star$) \tablenotemark{\ddag}} & \colhead{R$_{co}$ (R$_\star$)} & \colhead{R$_{trunc}$ (R$_\star$)}
  }
  \startdata
  6  	&	1.7	&	5830	&	17	&	2.86	&	4.04	&	3.56	&	0.28	&	0.04	&	0.04	\\
21  	&	2.5	&	5250	&	3.9	&	1.48	&	2.39	&	5.83	&	0.13	&	0.04	&	0.04	\\
41  	&	2.8	&	4060	&	0.79	&	0.55	&	1.8	&	5.76	&	0.06	&	0.03	&	0.03	\\
61 	&	30	&	3955	&	0.54	&	0.55	&	1.56	&	4.53	&	0.05	&	0.15	&	0.15	\\
140 	&	12	&	3379	&	0.13	&	0.28	&	1.05	&	3.41	&	0.02	&	0.07	&	0.07	\\
182 	&	2.7	&	3234	&	0.15	&	0.2	&	1.23	&	3.43	&	0.03	&	0.02	&	0.02	\\
\enddata
  \tablecomments{Derived stellar properties of the discarded sample. Refer to Table \ref{derivedtable} for notes.}
\label{discardedtable}
\tablenotetext{\dag \ddag}{Refer to Table \ref{derivedtable} for notes on masses, radii, and sublimation radii of our sample.}
\end{deluxetable}

\end{document}